\documentclass[12pt,a4paper]{article}

\usepackage[british]{babel}

\usepackage[a4paper,top=2cm,bottom=2cm,left=2.5cm,right=2.5cm,marginparwidth=1.75cm]{geometry}

\usepackage[natbibapa]{apacite}
\usepackage{amsmath}
\usepackage{graphicx}
\usepackage{ragged2e}
\usepackage[hyphens]{url}
\usepackage{etoolbox}
\usepackage[colorlinks=true, allcolors=blue]{hyperref}
\usepackage[hyphenbreaks]{breakurl}
\usepackage[title]{appendix}
\usepackage{mathrsfs}
\usepackage{amsfonts}
\usepackage{booktabs} 
\usepackage{caption}  
\usepackage{threeparttable} 
\usepackage{algorithm}
\usepackage{algorithmicx}
\usepackage{algpseudocode}
\usepackage{listings}
\usepackage{enumitem}
\usepackage{chngcntr}
\usepackage{booktabs}
\usepackage{lipsum}
\usepackage{subcaption}
\usepackage{authblk}
\usepackage[T1]{fontenc}    
\usepackage{csquotes}       
\usepackage{diagbox}
\usepackage{comment}

\usepackage{subcaption}
\usepackage[justification=raggedright, singlelinecheck=false]{caption}
\usepackage{caption}
\usepackage{helvet}  
\usepackage{lmodern}  
\usepackage{float}

\usepackage{setspace}
\usepackage{titlesec}
\titleformat{\section} 
  {\normalfont\Large\bfseries}{\thesection.}{1em}{}
  
\newcommand{\subsubsubsection}[1]{%
  \vspace{\baselineskip}
  \noindent\textbf{#1\\}\quad
}
\usepackage{float}   
\usepackage{caption} 


\title{The Impact of AI on the Cyber Offense-Defense Balance and the Character of Cyber Conflict}

\author[1]{Andrew J. Lohn}

\affil[1]{\small Center for Security and Emerging Technology, Georgetown University}
\affil[*]{\small \texttt{drew.lohn@georgetown.edu}}

\date{}  

\begin{document}
\maketitle

\begin{abstract}
Unlike other domains of conflict, and unlike other fields with high anticipated risk from AI, the cyber domain is intrinsically digital with a tight feedback loop between AI training and cyber application. Cyber may have some of the largest and earliest impacts from AI, so it is important to understand how the cyber domain may change as AI continues to advance. Our approach reviewed the literature, collecting nine arguments that have been proposed for offensive advantage in cyber conflict and nine proposed arguments for defensive advantage. We include an additional forty-eight arguments that have been proposed to give cyber conflict and competition its character as collected separately by Healey, Jervis, and Nandrajog. We then consider how each of those arguments and propositions might change with varying degrees of AI advancement. We find that the cyber domain is too multifaceted for a single answer to whether AI will enhance offense or defense broadly. AI will improve some aspects, hinder others, and leave some aspects unchanged. We collect and present forty-four ways that we expect AI to impact the cyber offense-defense balance and the character of cyber conflict and competition.  
 
\end{abstract}

\textbf{Keywords}: AI, cyber, offense-defense balance, conflict, crisis

\tableofcontents


\section{Introduction}
Of all the domains that AI is likely to affect, cybersecurity and cyber conflict may be the most significant \citep{gallupCyber, pewCyber}. As a digital domain, cyber is naturally amenable to incorporating AI advances. Cyber’s digital nature also benefits AI advancement in that useful data can be easily generated through normal operations, directed experiments, or through autonomous self-play. While cyber may not usually be as directly disastrous as risks from some other domains such nuclear war or biological pathogen development, cyberattacks can, in principle, lead to disasters in almost all other domains \citep{smith2021wmd,radoini2021wmd}.

As AI continues to progress and continues to be incorporated into nearly every domain, any clues about how those domains will change become more valuable. Whether the changes are for better or worse, knowing that they may come provides time for society to adjust and prepare. Although predictions about the future of technology are highly uncertain, considering possible futures allows policymakers, regulators, and society as a whole to make decisions that lead to preferable futures that are less risky.

In this article we try to take a comprehensive view of how AI may affect the character of cybersecurity and cyber conflict and ask whether AI will benefit offense or defense. The question of whether cyber has historically been offensively or defensively biased has been debated for decades. While scholars appear to lean toward offense, there is no consensus \citep{slayton2017balance, smythe2020cult}. This report does not try to resolve that historical debate, nor does it attempt to resolve whether cyber will be offensively or defensively biased by advances in AI in the future. We only try to raise arguments for how AI may be expected to make cyber more or less offensively or defensively biased than its current state. Regardless of whether cyber is offensively biased now or in the future, it is worthwhile to know whether it is expected to be more offensively or defensively biased in the future and to find ways to drive the balance further in favor of defense \citep{cybertask2017build,alperovitch2020understand}.

Methodologically, we conducted a review of the cyber offense-defense literature and engaged with several of the field’s prominent authors. The review’s purpose was to extract the arguments that have been made in favor of offense or defensive biases in cyber. Once the arguments were collected, we considered AI’s effect on each argument to evaluate whether and how AI might drive cyber to be more offensively or defensively biased.

It is possible, or even likely, that our literature review unintentionally omitted aspects of the cyber offense-defense balance. While not comprehensive, we expect that it does capture the most prominent arguments. The search also managed to uncover a paper that did not exist at the outset of this project and that is far more expansive than all others we found although it is not directly aimed at the question of cyber's offense-defense balance \citep{healeyCharacter}. The paper by Jason Healey, Divyam Nandrajog, and the late Robert Jervis, who is credited with initiating the theory of offense-defense balance, \citep{jervis1978cooperation} collects propostions for what gives cyber conflict and competition its character. It too falls short of being a perfectly complete accounting of the character of cyber conflict, but it is a worthy enough attempt that we consider each of their 48 elements as well.

Following this introduction, the report continues by describing offensive and defensive bias in cyber and the effect of being offensively or defensively biased. The third section of the report describes the various threat actors as well as the various targets or defenders. The fourth section describes the possible degrees of AI advancement: status quo, reliable and independent, expert, hard limits, and limit-breaking. The fifth section describes the potential to control access to AI advancement. The sixth section outlines various possible strengths and weaknesses that were used as prompts during brainstorming the ways that AI might affect aspects of the cyber offense-defense balance and the character of cyber conflict. Sections seven and eight contain assessments of AI's expected impact, cyber’s offense-defense balance, and the characteristics of cyber conflict respectively. Section nine aggregates the lines of reasoning from sections seven and eight for ways that AI might impact the cyber domain. And section ten concludes.

Each of the elements evaluated in sections seven and eight deserve their own thorough report. The evaluations done here should be considered preliminary and uncertain. They are an attempt to achieve a big-picture view of AI’s effect on cyber’s offense-defense balance and the character of cyber conflict. We anticipate conducting more thorough investigations of many of the individual elements identified here, and we invite others in the research community to conduct their own assessments. It is possible that some of the conclusions within these individual elements will change on further investigation or as time removes uncertainty and ignorance from the assessments. We hope that this report provides a sufficiently overarching view of AI’s effect on the cyber domain to prepare decision-makers and defenders while also providing a corpus of important questions for cyber and AI researchers to investigate.

\section{The Offense-Defense Balance}
The offense-defense balance is a more general concept in international relations than its cyber context. It was originally proposed as a way to assess or affect the likelihood of conflict \citep{jervis1978cooperation}. It has been frequently revisited in the more narrow cyber domain, and more recently discussions have considered AI’s influence on both the general offense-defense balance and the cyber-specific offense-defense balance \citep{corsi2024considerations,tang2024implications}. If technologies or approaches provide more benefit to defense than offense, then the incentive for either side to attack is reduced, and vice versa \citep{garfinkelODScaling}. In theory, the balance provides a measurement of stability or instability.

In a general military context, defense has historically been favored. As a rule of thumb, an invading force needs to be three times larger than the defending force \citep{mearsheimer1989assessing}. In the cyber domain, it is common to believe the opposite, that offense has the advantage \citep{healey2021offense, ispi2013offense, dod2011offense}. That is not a consensus, and it is certainly not always true, just as attackers do not always require three times as many soldiers as defenders in a ground invasion.

The offense-defense balance can refer to many outcomes beyond stability and instability or the propensity to attack. For example, it can be framed in terms of monetary expense. Defensive spending outpacing offensive spending has been said to imply that the balance favors offense, although offense can be expensive and cost estimates vary widely \citep{greenfield2023cybersecurity}.\footnote{An author of this report previously showed cost estimates for offensive operations that vary by six orders of magnitude in Box 4.1 of: V.A. Greenfield, J.W. Welburn, K. Schwindt, D. Ish, A.J. Lohn, and G.S. Hartnett, “Cybersecurity and Supply Chain Risk Management Are not Simply Additive,” RAND Report A532-1 (2023).}

The offense-defense balance can also refer to the extent of damage. In a monetary sense, damages have been substantial. In a military sense, the effects have been more muted. Many argue that cyber effects have historically been small and reversible although that may downplay cyber's value in persistently collecting valuable information. Cyber’s somewhat limited role in the Russian invasion of Ukraine and their preference for conventional domains lends some credence to claims that cyber offers less to offense than previously believed \citep{terajima2022ukraine}.

Alternatively, the offense-defense balance can be framed in terms of vulnerability or coercion. An aggressor could increase influence even while reducing risk of attack. For instance, the low barrier to entry in cyber may allow a smaller less advanced nation to threaten the many digital targets of a more advanced nation. The likelihood or effect of those attacks may decrease as cyber acts as a “pressure release valve” even as cyber increases the aggressor’s influence \citep{healey2020escalation}. Still separately, the offense-defense balance can refer to difficulty, inconvenience or efficiency. The frustration of remembering thirty passwords and working in air-gapped facilities are the consequences of an offense-defense balance that leans more toward offense than would be preferred.

Looking across the possible outcomes of interest, it is probably not feasible or even desirable to define a single offense-defense balance. Even if it is possible to define offense-defense balance in a way that satisfies its many different reasonable interpretations and implications, it is probably not so useful to assess whether the balance favors offense or defense. Regardless of whether the balance favors offense or defense, the goal remains to develop technology and policy that pushes the balance to further favor defense. In adding the cyber domain to previous domains of war: air, land, and sea, “The more strategically useful  question  is  whether  and  how  the  advent  of  cyber  capabilities  has  shifted  the  offense-defense  balance  between  states  more  broadly,” \citep{huntley2024offense}. 

In adding AI to the cyber domain, the strategically useful question is whether and how AI will shift the offense-defense balance. That shift is not simply among nation-states. Nation-states are important attackers, targets, and defenders, but they are not alone. The next section describes the threat actors, targets, and defenders that should be considered in the context of AI’s impact on the cyber domain.

\section{Types of Attackers and Defenders}
The concept that a balance between offense and defense determines the aggressiveness of rivals originated in the field of international relations outside the cyber context.\citep{jervis1978cooperation} As a result, the literature has focused more on nation-on-nation conflict, but there are many other possible attackers and defenders. Some work has broadened to include varieties of attackers and targets \citep{healey2021understanding}. Those can include criminals, who may or may not be protected by international borders and who might ransomware schools, hospitals, or oil pipelines. Additionally, cyber novices might use readily available hacking tools to disrupt servers worldwide. And nation-state hackers might target individual corporations to provide economic advantages to their own companies.

AI’s impact on these various attackers and defenders will be different, so we consider several categories of each to help separate those different impacts. The divisions are not crisp; a single organization can fit partly in two or more categories. 

\subsection{Cyber threat actors}
The Canadian government summarizes the categories of threat actors as follows: Nation State, Cybercriminals, Hacktivists, Terrorist Groups, Thrill Seekers, Insider Threats \citep{canada2022introduction}. IBM uses the same groupings, \citep{ibm2023what} and we will here as well.

\subsubsection*{Nation-states}
Nation-states have some of the most advanced and secretive capabilities in the world. They have large teams with significant infrastructure such as deniable networks for distributing attacks and command and control throughout their operations. They are the only threat actors that can claim legitimate use of offensive cyber. Their objectives for using their cyber offensives vary, and does their susceptibility to diplomatic pressure.

\subsubsection*{Cybercriminals}
Many of the most disruptive and problematic attacks have been conducted by criminals, usually for financial gain. Common examples are ransomware or other forms of extortion. These criminals are often located in countries that protect them from retribution from their foreign victims and may be directed or influenced by the host nation, sometimes blurring the distinction between criminal and nation-state \citep{cisa2022russia}. Criminal organizations can be relatively sizable and skilled, sometimes even composed of moonlighting nation-state hackers.

\subsubsection*{Cyberterrorists}
Cyberterrorists are motivated by disruption or destruction. Rogue nations can blur this category with the nation-state category. Cyber terrorists can potentially be relatively well-funded but have historically been relatively low in sophistication and technical ability.

\subsubsection*{Hacktivists}
Hacktivists are motivated by a political or social ideology. Their attacks are less often destructive than can be expected from nation-states or cyber criminals. Typical attacks are website defacements or leaks of incriminating data. Hacktivists tend to be less well-funded and have less technical capability and sophistication.

\subsubsection*{Thrill seekers}
Thrill seekers hack for curiosity or entertainment. This is the category of most script-kiddies who have little technical sophistication but can access publicly available hacking tools. Thrill seekers can also include grey hat hackers who may cause accidental damage while probing unknowing targets.

\subsubsection*{Insider threats}
Insiders can be particularly damaging with lower levels of technical capability because of their level of access. They can steal data or sabotage systems or codebases. Insiders may be disgruntled individuals or may be recruited by a nation-state or ideologically motivated hacktivists or terrorists. Insiders can also be well-intentioned but make damaging errors.

\subsection{Defenders or targets of attack}
A defender and a target are not necessarily the same entity. For instance, defense can be partly contracted out. A targeted organization might have antivirus developed by one company and DDoS protection offered by a different company. Additionally, defense happens across the entire digital ecosystem. Software developers follow security best practices, do defensive testing, and issue defensive patches. Even disaster recovery organizations offer improved resilience that is an important aspect of cyber defense.

\subsubsection*{National government}
The departments and agencies that make up a government are common cyber targets. Those could be elements of the Department of Defense or the National Labs developing or managing the nuclear stockpile. They could be the Office of Personnel Management that holds sensitive information about government employees or the Internal Revenue Service that maintains records for all citizens. And they could be organizations that provide domestic services such as the Federal Emergency Management Agency or Medicare.

Governments help defense with information sharing, standard setting, and many other services including network monitoring, and penetration testing, and tool development. National governments also play unique roles in defense with their authority and capacity to penetrate rival networks. They provide warnings about adversary capabilities and intent. They also perform digital operations that can disrupt attacks as they occur or prevent anticipated attacks. 

\subsubsection*{Critical infrastructure}
In the U.S., critical infrastructure is largely managed by State, Local, Tribal, and Territorial governments rather than national governments. There are sixteen sectors of critical infrastructure, including chemical, water, nuclear, transportation, and others \citep{cisacritical}. These systems are complex, varied, and numerous, making it difficult to defend them all well. Critical infrastructure is also an impactful target for cyber attacks, making these systems especially important to defend well. 

\subsubsection*{Large organizations}
Large organizations are typically well-resourced to defend themselves but have often been the targets of sophisticated and enduring attacks. These organizations can include defense contractors, major banks, and large technology firms. The various sectors may be targeted for different reasons such as military advantage, economic disruption, or intellectual property theft, but they have some defensive similarities. They all have the ability to spend significantly to maintain teams of skilled digital defenders who work around the clock monitoring for and responding to possible security threats and breaches.

Some of these organizations contribute to the defense of other targets. They develop or manage the software and digital infrastructure that cyberwarfare occurs within. Much of the work of preventing and thwarting attacks falls to those companies as they monitor their clients, repair vulnerabilities, and share information.

\subsubsection*{Small and medium sized organizations}
Smaller organizations cannot afford large teams of defenders nor the most talented defenders. These organizations can carry substantial risk nonetheless, and for example, make up much of the critical infrastructure sectors. Small organizations can also be critical nodes in a supply chain, disrupting large sectors when they stumble, or providing a soft entry point for attackers to reach higher-profile targets.

\subsubsection*{Individuals}
The nation and the world are made up of individuals who can each be targeted. Individuals can have money stolen from crypto wallets or potentially from banks. They can be targets of disinformation or theft. Individuals can be targeted en masse as in election interference campaigns and the Equifax hack, or they can be targeted individually as in Nigerian Prince scams.

Individuals are responsible for their own digital defenses to some degree. Installing antivirus and using two-factor authentication are common defensive steps that individuals take. People choose encrypted messaging apps and avoid clicking on suspicious emails or text messages. Individuals rarely conduct sophisticated scans but may employ autonomous defensive tools that run unnoticed in the background.

\section{Levels of AI Improvement}
AI may advance to varying degrees or may pass through various levels of advancement as progress continues. There are too many variables and uncertainties to confidently predict technological progress and sociological responses to that progress. Some of those variables are degree of originality, independence, reliability, speed, and scalability. These variables are interrelated. For example, speed or scale can imply some independence in that humans do not have time to assess decisions. Reliability can also result in independence if less supervision is needed, which can then further enable scalability.

In this section, we describe four levels of AI advancement: status-quo, reliable and independent, expert, hard limits. These four levels are meant to typify several of the variables or uncertainties related to both technical and sociological aspects of AI may progress. Not all of those technical or social aspects need to advance together. For instance, the future may have effective vulnerability discovery which requires expert-level originality while also having unreliable code generators that failed to advance from status-quo to the reliable and independent level.

\subsection{Status quo}
Computers have long been superhuman in various regards such as speed and memory but inferior to humans in creativity. AI is already allowing humans to improve their efficiency at nearly all tasks and will continue to do so even without further technological advancement. Those efficiency gains will occur as AI's use increases and as it becomes better integrated into cybersecurity tools and tasks. But today, AI is often too unreliable to be given full autonomy or to substantially uplift low-skill individuals. AI is error-prone and vulnerable to corruption or manipulation \citep{lohn2020primer}. That unreliability limits its independence and scalability. 

At the status quo level, AI can help accelerate cyber practitioners. It has started to help experts discover vulnerabilities and to either help defenders write patches or attackers craft exploits. AI helps programmers produce more code per hour, and tools such as Security Copilot help defenders manage their daily deluge of alerts of possible intrusions \citep{hulme2024state}.

It is possible that progress will slow and not advance much beyond these limited and fallible assistants. AI progress is running into various computing and financing challenges \citep{lohn2023scaling}. AI might also be running into data limits and is approaching the entirety of human-generated data across the internet’s data \citep{villalobos2024run}.

It may be possible to overcome data limitations by creating synthetic datasets, especially in digital domains such as cyber, so advances beyond status quo might be possible even without further advances in AI algorithms or training \citep{lohn2023autonomous}. But it is not clear how easy that data is to create, how useful that data would be, and how many aspects of cybersecurity can benefit from it. For instance it can be difficult for AI training to keep pace with adapting attacker tactics because there is always less data about newly-developed tactics or tactics that have not yet been identified by defenders \citep{moore2019valuing}. This has been a challenge for previous attempts to use AI for cybersecurity. If techniques such as inference scaling, chain-of-thought reasoning, and synthetic data generation do not deliver on their promise, then aspects of AI for cyber could stagnate at the status quo.

\subsection{Reliable and independent}
AI is currently limited because of its high error rates. AI typically requires step-by-step guidance and human review of all important outputs. If AI was reliable and independent then it could scale far beyond the limits of human guidance and supervision. It could also be allowed to operate faster than humans can supervise. 

This level of competence may be easier to achieve for common aspects of cyber that have a lot of data from human practitioners than for rare aspects that require the type of creativity or originality that typically comes from human experts. Common aspects of cybersecurity also often have specific guidance or instructions to follow such as in best practice documents or standards. Following common and well-tested instructions may be a more achievable goal than demonstrating ingenuity.

AI models and systems could be more reliable than humans but still be orders of magnitude less reliable than the components that are tyically certified for critical systems \citep{lohn2020brittleness}. New certification processes and standards might be needed to use even reliable AI components. They are also notoriously vulnerable to intentional manipulation or cooption \citep{lohn2020primer}. Despite these shortcomings, even potentially vulnerable AI may provide security beneifts in a cyber context where humans are often considered the weakest link \citep{schneier2018artificial}.

\subsection{Expert}
Human experts require little guidance and instruction. With only high-level instruction and intent, they can decide on a course of action and execute those actions with few errors. The ability to craft an unspecified course of action would provide AI with more agency. Human experts can also use their creativity to perform individual tasks that are rare and have little data to draw from. AI experts would be able to perform these individual tasks such as vulnerability discovery or use stealthy attack tactics as effectively as the best humans.

Expertise is a rare and distinguishing trait, so scaling expertise can have different effects than scaling basic competence. It could substantially level the playing field for defenders, nations, and threat actors. Small organizations could have access to creative imaginative and versatile defenders while a wider range of threat actors could have access to advanced persistent threat levels of expertise. That does not necessarily mean that all threats will be at a high degree of sophistication. Simple attacks have often proven effective and defensive effectiveness would need to improve to change that. 

The AI systems themselves could also remain insecure against corruption or manipulation. Historically, AI models have grown more manipulable with increasing capability \citep{ilyas2019bugsfeatures}. Human experts can be tricked, and might be easier to manipulate if, like AI systems, the manipulation attempts coud be tested billions of times on their target or one like it before being deployed. More capability in an AI model also typically provides more input space for the attack to hide and more internal calculations for the attacker to take advantage of.

\subsection{Hard limits}
AI may improve beyond the ability of even the best experts. This scenario is more plausible in domains where training data does not need to rely on human generation and in domains that are too varied and complex for a human to know them in their entirety. In domains like these, an AI may benefit from seeing more data than any human ever has and from being able to make connections between a larger number of sub-fields than any individual human has ever understood at one time.

Cybersecurity is such a domain. Reverse engineering, network architecture, compiler design, anomaly detection, and web interfaces are all related but distinct sub-fields. No human is an expert in all of cyber’s sub-fields. And being a primarily digital field, it is possible to autonomously create or collect mountains of data that no human ever needs to contribute to directly. Like teaching itself to play Chess or Go, an AI agent could simulate cyber battles against itself, learning more than any human could teach it \citep{lohn2023autonomous}.

But that does not mean that AI would be omnipotent. An extremely advanced intelligence would still face hard limits on what is possible. Some things can be mathematically proven to be impossible or proven to require more resources than is feasible. For example, brute force cracking a random password requires an average number of guesses and cannot be done faster than the available compute power allows. Or even more simply, attackers cannot exfiltrate information faster than the bandwidth that is provided, and they cannot store more information than the memory provided.

For more complex systems, there is software that can be formally proven to be secure. Those proofs rely on assumptions such as that the underlying hardware is reliable, but it is impossible for even a superintelligent AI to break the software directly. That lets defenders focus on the smaller job of protecting the assumptions rather than the software itself. Separately, most of the encryption in use today is not provably secure, experts just believe it to be difficult to break. A sufficiently intelligent AI could break current encryption unless we switch to provably secure approaches. From a defensive standpoint, perhaps such an intelligence may be able to help design provably secure implementations.

\subsection{Breaking limits}
Some theorize that AI can become so intelligent that it could exceed what we believe are hard limits. We consider this level of advancement out of scope for the current report. Perhaps AI could reason about undecidable problems such as the halting problem or Rice's Theorem \citep{kozen1977rice}. Perhaps it could break Shannon’s laws about entropy limits or signal-to-noise ratio for storing or concealing information \citep{shannonNoise}. Perhaps it could predict the behavior of chaotic systems. 

It does not seem likely that an AI would be able to break these limits. It is more likely that a sufficiently intelligent AI would create unanticipated ways to invalidate assumptions about the limit or the system that relies on it rather than break the limits themselves. Either way, it is hard to reason about an intelligence that, by definition, has no limits.

\section{Access to Advances}
In addition to the degree of AI advancement, it also matters how that advancement comes about. It matters who has access to it, how expensive it is, how memory-intensive it is, how quickly it operates, and how much difference there is between the best, the fastest, the cheapest, and the rest. 

\subsection{Controlled access to capability}
Technologies can be difficult to develop and sometimes it can be protected to maintain market share, protect intellectual property, or by export control restrictions for instance. AI technologies will certainly proliferate, but it is possible that the most capable AI models or systems could be contained within one or a few organizations that restrict access. There are many ways to define capable that must be considered. It is usually thought of in terms of accuracy, but other traits such as creativity, speed, or efficiency could be more important depending on the application. A “smarter” model may be bested by a dumber fast one if the contest is a race. An inexpensive dumb model may prevail if the contest is decided by the volume of attacks and defenses launched. And even a model or agent that is better in every way may provide no benefit if the task is simple enough that an inferior model is sufficient. 

This controlled access scenario could arise in several ways. The models or agents may be expensive to train and run, feasible only for a few well-funded organizations. Alternatively, some secret or patented conceptual advance such as a new algorithm or data generation procedure may provide the differentiating breakthrough. A new computer chip design or manufacturing technique could also provide an advantage that is difficult to replicate. 

Alternatively, controlled access to AI for cyber applications may not require an oligopoly on AI more generally. Organizations could develop a set of tools or data that are necessary for AI models or agents to be effective in cybersecurity applications. In that case, one can imagine governments with their classified data and tooling sustaining a lead in AI for cyber, especially offensive cyber, without needing to lead in AI more generally.

\subsection{Limited control of access to capability}
Even if only a few organizations are able to provide capable AI services, they may have limited ability to constrain or influence how its products and services are used. For instance, the provider could choose to accept or reject clients, choose which requests to accept or reject, and can log user activities for forensic investigation \citep{ji2023controllingLLMs}. But a large number of users and requests may make those actions practically infeasible.

The providers may also be limited if users can turn to less scrupulous service providers. Those alternative providers may be hosts from an adversary nation or providers with more permissive policies and ideologies. In an oligopoly where providers disagree about how to constrain or manage services, there may still be room for some agreement. For instance, even adversary nations may agree to restrict terrorist users and activities or to restrict certain types of criminal activities. 

\subsection{Proliferated models and agents}
An alternative possibility is that AI capabilities for cyber proliferate widely such that it is difficult to restrict or observe illegitimate users or uses of AI. That could happen if the expense and expertise needed to create or reproduce competitive models or agents is relatively low. It could also happen if one or more developers choose to release their models or agents openly. And without openly releasing models, providers may offer products that are closed source but run on the user's devices or servers. It may even be possible to do so in ways that prevent them from being inspected or altered \citep{brundage2020trustworthy}. In that case, there may be some ways to restrict who can access their products or services but little opportunity to restrict or monitor how they are used.

\section{AI Capabilities to Consider During Evaluation}
AI promises to improve performance in many ways including lowering costs, increasing speed, and perhaps reducing error rates \citep{pratt2024advantages}. Experts also anticipate possible downsides such as bias, security vulnerabilities, and perhaps increased error rates \citep{rashid2024ai}. This section lists some of these proposed strengths and weaknesses. The strengths and weaknesses in this section sometimes overlap with each other or even contradict each other. We do not make any assertions about which will be true. These lists were used as primers for ways that AI may change cybersecurity during the structured brainstorming described in the following sections. 

\subsection{Strengths}
\subsubsection*{Speed}
Computers already act quickly. Many of the aspects of both offensive and defensive cybersecurity are already highly automated and will not be accelerated by AI. Further, some activities, especially offensive activities, are performed slowly on purpose for reasons such as to avoid being flagged for suspicious behavior. But where human analysts or operators are the bottleneck, AI may be able to further accelerate this already high-paced domain. 

\subsubsection*{Availability}
Availability refers primarily to the ability for AI to operate 24/7 without growing tired or less vigilant. That advantage is more clear when comparing to individual humans than to teams of humans who can alternate shifts to work around the clock. 

\subsubsection*{Access to expertise}
AI provides expertise or experience that may otherwise be hard to come-by. It may be knowledge of many programming languages, knowing the word-for-word contents of a 500-page cybersecurity standard, or the best practices for configuring a firewall. This may be especially applicable to organizations that cannot afford top human talent.

\subsubsection*{Scalability}
Digital systems, including AI can usually be easily replicated arbitrarily many times. In addition to the number of AI systems, they may also increase the number or efficiency of operators and provide an increase in the effective scale of humans. They can also drive costs down to increase the number of efforts that a defender or attacker can afford. In addition to the number of systems, each individual system may be able to increase the scale of a task or operations such as by analyzing larger volumes of data. 

\subsubsection*{Complexity and silo breaking}
AI can make sense of data or scenarios that are more complex than humans can analyze or understand. It can also consider data from a wider range of sources, allowing it to make connections that humans would struggle to make if the key pieces of information are distributed to separate people.

\subsubsection*{Improved accuracy}
AI’s accuracy exceeds that of humans in an increasing number of domains \citep{salem2024advancing}. On the other hand, AI is prone to confabulations \citep{greaves2023confabulations}. Even where AI may be more reliable than humans, it can remain orders of magnitude less reliable than the components typically certified for critical systems \citep{lohn2020brittleness}. Systems that have been designed to accommodate fallible humans may need to be redesigned and reevaluated to be viable with fallible AI components.

\subsubsection*{Mundane tasks}
Some tasks, such as code inspection and refactoring or monitoring repetitive signals and alerts, could be easily conducted by humans but those tasks are not interesting. The humans who could perform those tasks choose to perform other tasks instead. AI systems have no preference against working on valuable but uninteresting tasks.

\subsection{Weaknesses}
\subsubsection*{Deskilling}
As AI performs more tasks that would otherwise be performed by humans, those humans may lose skills that they would otherwise have \citep{lee2025impact}. A familiar example is humans losing the ability to use a map or navigate by the stars \citep{navigatebystars}. In addition to human deskilling, systems may be designed in ways that preclude human operations. Autonomous vehicles may not have steering wheels, stores no longer sell maps or sextants, and AI-enabled industrial control systems may not build in human interfaces. These changes could make AI-enabled systems more vulnerable to attack and impede resilience and recovery.

\subsubsection*{Bias}
AI systems reflect biases in their training data. Bias in AI is usually discussed in the context of human rights and equity, but in cybersecurity it could also refer to systematic trends in the data or outcomes. For example, a biased cyber system may priortize certain types of vulnerabilities over others. As another example, there are legitimate differences in preference between availability and security or privacy. An AI system may act in ways that prioritize ease of use over privacy or vice versa. 

Even if an AI system is less biased than humans, AI bias can be problematic if a single bias is systematically propagated at scale where many separate humans would have led to a more diverse set of biases and more resilience.

\subsubsection*{Security}
AI models and systems are notoriously vulnerable to a wide range of attacks. It is likely that both offensive and defensive AI agents will be vulnerable to subversion or even co-option as well as corruption such as through poisoning attacks \citep{lohn2020primer}.

\subsubsection*{Transparency}
The inner workings of AI systems are uninterpretable to humans \citep{vonEschenbach2021transparency}. The inner workings of humans are also uninterpretable to humans, but shared experience allows humans to relate to each other.

In addition to the transparency of the AI models or systems themselves, transparency can also refer to the products that are created by either AI or humans. For example, humans write code in human-readable languages but AI systems might write code or design systems in ways that are efficient for machines but difficult for humans to interpret. At the same time, AI has excelled at summarizing or explaining complex ideas or systems and may help make opaque systems more understandable for humans.

\subsubsection*{Alignment to Human Preferences}
AI systems may struggle to understand the human experience, human emotions, human preferences, and human ethics. It may do a poor job of interpreting human guidance and intent. It may also do a poor job of anticipating human responses to its actions or choices. But humans can also find it difficult to interpret the intent, values, or likely reactions of others, so it may be possible in some cases for AI to match or exceed human ability in these domains.

\subsubsection*{Confabulations}
AI systems regularly make false pronouncements confidently \citep{aljamaan2024hallucinations, magesh2024hallucinations}. They can make errors that humans would not and in ways that can be difficult to detect.

\section{Asymmetries Between Cyber Offense and Defense}
\label{asymmetriesSection}
There have been many arguments put forward for either offensive or defensive advantage in the cyber domain. We do not try to defend or refute these arguments, nor do we try to weigh them to determine whether offense or defense has an overall advantage. We simply reviewed the literature to collect these arguments and consider the ways that advances in AI are likely to strengthen or weaken those arguments. We collected the nine arguments that favor defense and nine arguments that favor offense listed in Table \ref{offense-defense-bullets}. In the rest of this section we describe how advances in AI might affect these arguments and shift the balance of the cyber domain more toward defense or toward offense.

\begin{center}
\begin{tabular}{ l l } 
 \textbf{Defensive Advantages} & \textbf{Offensive Advantages} \\ 
 $\bullet$ Defenders determine the digital terrain & $\bullet$ Attackers only need one success  \\ 
 $\bullet$ Defense can be resilient & $\bullet$ Attackers choose when to strike  \\ 
 $\bullet$ Defenders can observe their networks & $\bullet$ Attackers choose who to strike \\ 
 $\bullet$ Defenders can provision local resources & $\bullet$ Attackers choose their goals \\ 
 $\bullet$ Attackers must penetrate all defenses & $\bullet$ Targets are easy to find \\ 
 $\bullet$ There are more defenders than attackers & $\bullet$ Defense requires reliability \\
 $\bullet$ Attack takes time & $\bullet$ Defense has more bureaucracy \\ 
 $\bullet$ Offense risks escalation & $\bullet$ Defense has systematic vulnerabilities \\ 
 $\bullet$ Attackers are illegitimate & $\bullet$ Defense has systemic vulnerabilities \\ 
\end{tabular}
\label{offense-defense-bullets}
\end{center}

\subsection{Defensive Asymmetries}
\subsubsection{Defenders determine the digital terrain}
In a land battle, defenders can dig trenches, lay mines, build walls around castles, and place artillery on hilltops. It gives them their rule-of-thumb three attackers to every one defender advantage \citep{mearsheimer1989assessing}. The digital domain is inherently artificial and reconfigurable, so defense’s ability to define the environment may be even more pronounced in cyber conflict.

Network connections, bandwidth, air gaps, firewalls, blacklists, password rules, access permissions, and countless other configurations are all set by defenders. Defense gets to decide on the requirements or functionality of the systems and services they implement and can choose not to use faulty or vulnerable software or systems. Defense also designs more global aspects of the digital environment such as the internet’s addressing system, encrypted communications, and even the physical infrastructure of cables and switches. 

\subsubsubsection{Speed}
Defenders can take the time to configure their defenses prior to going live and exposing them to attackers. But they could be pressured to deliver on business timelines or to adapt quickly during crisis or conflict. Those near real time adjustments might mean that an attacker who had a foothold might find it unreachable in precisely the moment they seek to use it. Or an attacker might find their malware or implants flagged by new search terms just as they intended to install them. 

Some of this is possible without AI, but for humans to match AI’s speed would require more preplanning. Humans could prepare a set of configurations that correspond to levels of heightened alert. Some of those changes could be automated and would occur just as fast under human command as AI command. Reliable and independent AI systems could be more adaptable than pre-planned procedures but some changes may be high stakes and unwise to delegate. Killing processes on a victim’s computer, wiping that device, or temporarily disabling services could all be risky defensive actions. And if the defensive agent can be tricked or manipulated by the attacker, then the defensive agent’s reconfigurations could cause more harm than good.

\subsubsubsection{Access to expertise}
It is not easy to successfully attack a target that is well-configured for defense, but it is also not easy to make a target well-configured for defense. Large companies with large teams of defensive experts configure defenses well and are hard to attack. Most organizations, including much of critical infrastructure, are less well-configured. Status quo levels of AI can likely help moderately-skilled defenders harden their systems. If expert level AI becomes widely available, then every organization could harden themselves similarly to the best defended large companies. Those advances would level the defensive playing field, substantially benefiting individuals, small-medium sized organizations, and critical infrastructure. Smaller organizations might even become harder targets than large organizations because of their smaller attack surface.

Against hard targets, attackers need to be more inventive about their intrusions, such as by using more zero-day vulnerabilities. Attackers are already trending toward increasing reliance on zero-days, possibly due to improving defensive preparations \citep{cisa2024top}. While it is difficult for attackers who rely on known tools and techniques to succeed against hard targets, the common wisdom is that the best attackers will breach any target if they have enough time. That implies that expert-level AI attackers would often succeed against even expert-level AI defenders. 

Nevertheless, attacker advantage at the expert level or beyond is not a foregone conclusion. It relies on there being many novel exploits or techniques for attackers to discover. It is possible that defending at AI scale will find and fix most of them, leaving even expert or superhuman AI attackers empty handed. Defenders may also design their systems in ways that are less vulnerable, even provably secure in some cases. If AI defenders can increase the fraction of systems that are provably secure, then even superhuman AI attackers would see their attack avenues decrease. Better design may allow defenders to progressively shrink their defensive tasks and focus on a smaller number of vulnerable paths.

\subsubsection{Defense can be resilient}
Defenders decide how much risk they are willing to accept. As examples, defense sets those limits by maintaining secure backups, manual modes of operation, and paper ballots for elections. It would be convenient to wire the nuclear arsenal straight to the open internet, and it would be useful for air traffic control to be able to remotely pilot non-responsive or erratic aircraft, but some things are too risky to accept even the potential for cyberattack. Instead, defenders set limits for the maximum amount of damage that cyberattack can cause \citep{gray2013making}.

\subsubsubsection{Scalability}
This resilience comes not from what the system \textit{can} do, but from what the system is \textit{prevented from doing}. AI can be used to help identify unacceptable risks which can then be avoided or managed. The second category of NIST’s Cybersecurity Framework is Identify, which is meant to understand the organization’s cybersecurity risks \citep{NISTCSF}. That can be difficult because of the potentially large inventories of digital assets in an organization and because of the large number of possible scenarios that could arise from misuse of any of them. 

AI may help alleviate this problem with its ability to take in vast amounts of data and assess many scenarios. AI can likely contribute somewhat to those efforts without much further improvement such as by helping to brainstorm hypothetical risks. Further AI progress could make AI more useful by increasing its accuracy and inventiveness. Superhuman intelligences may be able to identify and eliminate risks that humans would never consider, especially in complex systems where the complete set of risks are not readily apparent.

\subsubsubsection{Deskilling}
The existence of humans in targeted systems and organizations is a form of resilience against cyber attacks. Although humans are often the weak link, they are not directly controllable by hackers. Replacing humans with digital entities threatens to make those systems more vulnerable, especially if AI systems remain more vulnerable to cooption and manipulation than other digital systems.

Aside from being built of vulnerable components, integrating AI may adversely affect recovery and resilience efforts. For example, autonomous vehicles without steering wheels would be difficult to revert to manual mode in the case of attack. AI integration could lead to a widespread loss of human-machine interfaces across all sectors, but even if interfaces remain, humans may lose the skill to use them.

\subsubsection{Defenders can observe their networks}
Defenders can instrument their networks to provide logs and scans. They can observe everything that goes into a network, any processes that are running on any of the endpoints, and the behavior of any of the users. There are typically too many endpoints and processes to observe everything, but defenders can choose to observe almost anything. And defenders do not need to worry about the stealthiness of their actions as long as those scans and monitors do not degrade performance too much. 

With enough time and effort, attackers can learn a lot about a network they study too. When Rob Joyce was head of NSA’s offensive cyber team, he said "We put the time in ...to know [that network] better than the people who designed it and the people who are securing it... You know the technologies you intended to use in that network. We know the technologies that are actually in use in that network," \citep{zetter2016nsa}. This may be especially true as people blur their personal and official devices, especially while working from home.

\subsubsubsection{Availability}
Scanning can already be done at scale, but the skill to implement the scans and to interpret them can be limited. Large organizations have security professionals working around the clock to monitor their networks but smaller organizations do not have that luxury without AI. AI that can independently and reliably investigate alerts and respond to them, or that can adapt those scans and logs to the threat environment, would be a substantial boon for defense.

\subsubsubsection{Complex data and silo breaking}
The logs, data, and alerts flooding human defenders can be overwhelming to manage. That data is typically truncated in order to be manageable, so potentially useful data is thrown out to avoid overwhelming human defenders. AI systems can view more data, and potentially make connections that humans would not make. AI has long been superhuman in its ability to ingest large amounts of data but not always in its ability to makes sense of that data. 

There may be hard limits on observability that could affect both attackers and defenders. The field of signal detection theory is filled with mathematical results describing the thresholds between what is detectable and undetectable. Many of those theories originated in cybersecurity or have been widely applied in cybersecurity to set hard limits for what defenders can observe and for how aggressive a stealthy attacker can be.\citep{ding2017survey}

\subsubsubsection{Security}
AI models are vulnerable to deceptive patterns and can be made to see things that are not there or to miss things that are \citep{lohn2020primer}. This effect is especially infamous from image processing models but is true in all categories of AI, including cybersecurity. AI may struggle to improve observability in the networks or systems it defends if attackers are able to obscure their behavior by taking advantage of this weakness. With further progress, AI systems may become more difficult to deceive, but historically, making AI more capable by increasing the complexity of its inputs and decision processes has increased the number of ways it can be deceived. The more advanced the AI, the more difficult it has been to secure.

\subsubsection{Defenders can provision local resources}
Defenders can provision as much memory, computational power, and bandwidth as they can afford. They can establish high-throughput connections to powerful computers that are dedicated to analyzing suspicious behavior or planning responses to attacker actions. Attackers on the other hand can either use the local resources of their victims or transmit data to and from a remote command and control server. In either case, they risk alerting defenders who can remove the attacker or cut off access in various ways including simply pulling the plug on the targeted systems if necessary.

Attackers may therefore be restricted to AI models or agents that use very low memory and compute, or limit the amount of information supplied to and received from remote models. For example, defenders could easily scan a gigabyte of software looking for vulnerabilities, but an attacker could neither scan that gigabyte nor send it to be scanned without risking detection. These restrictions on offense could give defense an advantage to select more capable AI systems and to run them more efficiently on appropriate hardware with better access to information. 

This discrepancy only applies to engagements that occur in the parts of the digital environment that defense controls. Some offensive activities such as vulnerability discovery in open source software can be done completely remotely without sending information to and from victim machines. There are elements of each stage of MITRE’s ATT\&CK Framework that would face these limitations, and there are elements of each stage that would not \citep{mitreattack}.

\subsubsubsection{Availability}
Defenders’ ability to provision resources means that, at least in some cases, defenders will have the ability to choose more capable models and agents than the attackers they face. The most resource-intensive models and agents will not be available to attackers for some tasks. 

That alone does not mean attackers are disadvantaged because attackers and defenders perform such different tasks. For example, establishing persistence in a running process is not a particularly similar job to identifying and terminating a process that is providing persistence. AI’s ability to help attackers despite their resource limitations will probably need to be evaluated on a task-by-task basis in order to determine whether offense or defense is advantaged by AI progress during digital engagements.

\subsubsection{Attackers must penetrate all defenses}
Defenders do not rely on any single defense. For example, attackers may need to communicate with a victim machine without triggering firewalls, install software without triggering antivirus, and transfer information or execute commands without triggering anomaly detection. A misstep at any one of these stages could foil the attack. 

There are cases where a single vulnerability enables a complete attack. This is more common in exposed services where users are granted direct access to core systems or data. An example is the heartbleed vulnerability which allowed attackers to access the private information of other users \citep{heartbleed2020}. It was common because websites that allow users to directly access their own information often also hold other people’s information on the same servers. Since the system was designed for users to access those servers to retrieve their own information, there were few defenses to overcome.

\subsubsubsection{Speed}
If the objective of defense in depth is to slow attackers enough for them to be detected and expelled then the speed of attack and defense matters \citep{lohn2019defense}. If AI accelerates defenders’ ability to detect and expel intruders more than it accelerates attackers’ ability to move through defenses then defense benefits more. But the opposite could be true. And even if the acceleration favors defense in each instance, an increased rate of attacks could lead to a larger number of successful attacks within a given period of time.

Defense has the ability to control speeds to some extent. Encrypting a harddrive takes time and can be made slower. Suspicious behaviors can be delayed with pauses like the lockouts that occur after too many incorrect password guesses. Suspicious or dangerous activity such as encryption, deleting the master boot record, or large file transfers could be delayed for human approval. These delays might interfere somewhat with productivity, but productivity can be sacrificed if security is prioritized.

\subsubsubsection{Scalability}
Alternatively, the objective of defense in depth can be to create a sufficient number of barriers so that an attacker will not penetrate all of them in a single campaign \citep{lohn2019defense}. In that case, the scale of AI may help to increase the number of defenses that a target can implement and monitor. But from the offensive standpoint, it may also increase the number of independent attackers the target faces. 

The ability to launch many different attacks in the same amount of time means that defenders must decrease the probability that any individual attack succeeds to keep the overall rate of intrusion unchanged. At the same time, if attackers increase the number of attempts, they increase their chances of being discovered. Those discoveries can help defenders track the attacks back to their source and interfere with the attacker's ability to launch attacks. Disrupting the attacker's infrastructure or operations can prevent future attacks as well as stop current ones.

\subsubsection{There are more defenders than attackers}
Nearly everyone needs to defend their systems, either by defending those systems themselves or by having a service conduct defense on their behalf. By comparison, there are relatively few attackers. Defenders use their ubiquity to their advantage by sharing information. A vulnerability or attack technique discovered in one system is quickly shared with other defenders, often in automated ways such as by updating YARA rules or antivirus signatures.

Even attackers contribute to collective defense, perhaps more than most defenders. Attackers are often targets for sophisticated espionage and intrusion campaigns such as from nation-states. The defensive services on those attackers’ machines are more likely than most to discover novel techniques that can then be shared among all defenders. Attackers also share their offensive techniques and tools with each other through various hacking forums or tool suites such as Metasploit. But that sharing also benefits defenders who can observe those forums or access those tool suites and adapt to any new disclosures. 

The ubiquity of defenders also creates a large market for defensive innovations, enabling economies of scale and creating competitive pressures that drive rapid progress. That defense is so expensive has been used to argue that offense is advantaged in cyberspace. However, defense’s greater budget may provide advantages in other measures of the offense-defense balance such as increasing the difficulty of causing damaging effects or in providing stability during crisis and conflict.

\subsubsubsection{Scale}
Reliable and independent AI agents could increase the number of defenders per target. AI could also increasing the number of attackers by lowering the barrier to entry and it could increase the number of agents per attacker. Advances in AI could also increase the number of targets and the attack surface of those targets if AI increases the rate of software development.

Defenders will probably continue to reap the benefits of information sharing as AI continues to scale. AI defenders will be able to share their discoveries and effective techniques among each other in a way that benefits all defenders. AI-enabled attackers would still suffer from the risk that sharing exploits and techniques would inform defenders. But there are cases where that could be less of an impediment for attackers. For example, hardware vulnerabilities can be difficult to patch, so the risk of informing defenders would be less pronounced \citep{lohn2018meltdown}.

\subsubsection{Attack takes time}
Attacks often progress through several stages such as those outlined in the “cyber kill chain” or Mitre’s ATT\&CK framework \citep{lockheedcyberkill,mitreattack}. Sometimes stages can be completed in parallel, but they often rely on each other in sequence, and stages may recur several times. For instance, reconnaissance may be required before initial access and then again for resource development. Resource development might be needed for initial access and again for privilege escalation or lateral movement. 

The long timelines for conducting an offensive campaign can be time consuming can help explain why nation states choose to preposition cyber capabilities in critical infrastructure during peacetime. Attackers may also choose to delay some stages of attack to avoid “burning” capabilities by giving defenders the opportunity to observe them and adapt. It has also been suggested that Russian cyber offensives were limited in the 2022 Ukraine invasion because their cyber attackers were given insufficient time to prepare \citep{lee2023russian}. 

\subsubsubsection{Speed}
Any acceleration of the attack timeline weakens this defensive advantage. Even if defense is accelerated more than offense, the prospect that offense could attempt their intrusion in days or weeks instead of months improves the offensive strategic position. They would be able to conduct their attacks with less lead time and give defenders less forewarning. A cyberattack that may have been too slow to prepare in time for a days-long ground invasion or a minutes or hours-long missile strike may become viable.

But accelerating the overall timeline of offensive campaigns is a difficult proposition because accelerating many, but not all domains, could have limited effect and because many aspects of cyber are already so fast. If any of the parts of a cyber campaign remain slow, then the entire campaign will remain slow.

Additionally, speed of attack may be less important for non-nation-state attackers who have fewer obvious reasons to pair cyberattacks with real-world timelines. That said, there are examples to the contrary such as opportunist cybercriminals who might need to match their attacks to various opportunities that present themselves such as stock market fluctuations, natural disasters, or leaks of exploits or effective offensive tactics.

\subsubsubsection{Scale}
If the target is not specific, then scale can provide a type of speed of its own. If the objective is only to create havoc in an industry or geography, then the ability to simultaneously attack the police station, the nearby dam, the electric grid, and the fire department increases the odds that at least one of them will be compromised on the required timeline. Similarly, if AI increases the scale of development in ways that increase the attack surface, then attackers may be able to compromise their targets more quickly as well. 

\subsubsection{Offense risks escalation}
There are bureaucratic rationales that disadvantage offense. For example, offensive actions risk diplomatic blowback or retaliation that could be digital, kinetic, or legal \citep{valeriano2019myth}. A victim nation-state could respond to a cyber attack on critical infrastructure by bombing the aggressor. Or criminal hackers could be arrested or fined for their actions. As a result, cyber attackers are often hamstrung to some degree. Even cyber criminals who are protected within foreign borders likely recognize that the leniency of their protectors and the restraint of their victims only extends so far and constrain their attacks accordingly. 

As a counterargument, this defensive advantage has not always prevented attackers from conducting damaging or widespread attacks. There have been wanton attacks such as the WannaCry and NotPetya worms \citep{ccdcoe2021wannacry}. There have been attacks on critical infrastructure such as the Colonial pipeline, the electric grid, and hospitals \citep{dni2024infrastructure}. And there have been attacks on nuclear infrastructure \citep{zetter2014unprecedented}. Cyber's reputation for low escalation may actually make it more likely to be used in cases where escalation is likely.

\subsubsubsection{Reduced error rates}
If offensive AI can reduce error rates then attackers may be willing to allow AI agents to conduct cyber attacks that they would not trust humans to conduct. That delegation would require AI that is at least reliable and independent because errors are rampant at the status quo level. That level of trust might even require superhuman levels of reliability rather than merely expert levels because many important operational, and even tactical, decisions are made by experts rather than being delegated even to reliable humans.

\subsubsubsection{Transparency}
The decision to use offensive cyber is based on trust that is not only determined by error rates, it can also rely on transparency. AI systems are notoriously opaque themselves, and their actions or decisions can be opaque as well. But if AI is able to explain its plans and actions better than humans, then it may be transparent enough to gain the trust of decision-makers. In that case, AI attackers might be allowed to conduct attacks that humans would not be permitted to conduct. That would require substantial advances beyond the status quo wherein AI is currently viewed as less transparent than humans.

Another aspect of transparency that can influence offensive decisionmaking and the risks of escalation relate to attribution. It is not clear whether advanced AI would help or hinder attribution. If attackers are less likely to be identified, then those may be at lower risk of escalation. If similar AI technologies are used by many attackers then attribution may be more difficult. It may also be easier for AI to conduct false flag operations. And if AI conducts more parts of an attack, there may be fewer clues from its human operators to use in attribution. But many clues such as infrastructure, motives, and intercepted communications would remain. AI may also help make sense of diverse sources of information to conduct attribution more accurately and rapidly. And if various attackers use their own unique versions of AI then there may be unique characteristics of those AI systems that could be used to identify attackers.

\subsubsubsection{Alignment to human preferences}
AI struggles, and may continue to struggle, with understanding how humans feel and how they will react to its actions. The ability to determine which actions would be escalatory is difficult for humans and can be expected to be more difficult for AI. That may be exacerbated by bias such as might occur if the training set is built from the attacker’s culture rather than the target’s culture. This shortcoming could make attackers reluctant to use AI agents or could increase the risk of inadvertent escalation if they do.

\subsubsection{Attackers are illegitimate}
Attack is permitted as fair game when it is limited to espionage conducted by nation-states, but it is outlawed in almost all other cases. Even in espionage, legitimate cyber attacks risk loss of international and domestic support. At subnational levels, companies, organizations, and individuals are very limited in their ability to legitimately use offensive techniques to defend themselves or to retaliate. 

This is more of a sociological issue than a technical one. AI is unlikely to directly change this dynamic. It is possible that social norms and expectations could change as a result of technological advancement. For instance, companies may be granted leniency to use more offensive techniques if defending against AI attacks were to become much more difficult without the ability to disrupt attacks nearer to their source rather than in the target’s environment. 

Over time, and with more attacks, society may also become numb to cyber attacks in a way that does not make them officially legitimate but increasingly tolerated. Alternatively, society could decide that enough is enough and harden their stance against cyber attacks that would previously have been tolerated.

\subsection{Offensive Asymmetries}
\subsubsection{Attackers Only Need One Success}
Attackers can attempt their attack campaigns repeatedly until they succeed. They must defeat or bypass the full set of defenses in that campaign, but they only need a single campaign to be successful to achieve many of their goals. Any individual failure costs them little unless alerting defenders to the attack reveals their techniques or infrastructure in ways that make subsequent campaigns more difficult. 

Attackers may need their successes to occur on a short timeline, so defenders may only need a few successes if those successes delay attackers until a crisis has resolved or until some data or code has been finalized and cannot be corrupted or stolen. Further, if the attacker’s goal is to persist within the target, they may need to succeed many times if they are iteratively discovered and expelled.

\subsubsubsection{Speed and Scale}
If AI enables more attacks and defenses to occur in the same amount of time, either by occurring faster or from a greater number of attackers, then defenders would need to increase their probability of defensive success to keep the intrusion rate constant.

\subsubsubsection{Access to Expertise}
Only needing a single successful campaign may imply that discovering new tactics, techniques, procedures (TTPs), and vulnerabilities is an easier problem for offense than for defense. Imagine an inferior offensive AI agent facing a superior defensive AI agent. Imagine that the defensive agent is able to identify one million likely attacker TTPs or vulnerabilities and mitigate them and that the inferior offensive agent only identifies one thousand TTPs or vulnerabilities to attempt. If any of those thousand are not in the defensive agent’s million then the attacker has found a viable approach.

That scenario describes two agents that each draw from a random pool of TTPs and vulnerabilities. An alternative scenario could be where a superior AI agent finds all the TTPs and vulnerabilities that an inferior agent finds and more. In that case, offense would require an equally capable agent. Further, a campaign may require several of these TTPs or vulnerabilities. In that case, attackers might be concerned that failed attempts would reveal their novel TTPs or vulnerabilities.

\subsubsection{Attackers Choose Who to Strike}
In some cases, attackers have a specific target in mind such as if ground forces intend to breach a specific building and they need the door locks to be disengaged. Often attackers have more flexibility in their targets. They can install ransomware on any individual or organization who might be able to pay. They can choose the hospital, school, or politician with the worst digital defenses. Or they can pursue whichever artillery unit downloaded their malicious app \citep{crowdstrike2016use}.

\subsubsubsection{Scale}
Aside from the scale of attacks, AI-enhanced digital development might increase the number of targets for attackers to choose from. A growing codebase and a growing number of digital systems and applications would increase the amount of defending to be done.

In terms of attacks themselves though, even the earliest cyber attacks could be automated and replicated at scale \citep{orman2003morris, nguyen2024fast}. Those automated attacks did not allow for much differentiation among targets. For example, a ransomware worm will encrypt whichever target it reaches, although it may have simple rules for deciding which victims to engage or disengage. For example, Russian criminal groups disengage in some countries or if the target has the Russian language keyboard installed \citep{krebsDisengage}. Reliable and independent AI offensive agents may allow attackers to move further beyond such simple rules for targeting and perhaps make more fine-grained decisions about whether to engage a target in the first place. 

That increased intelligence could actually decrease the scale of attacks, or decrease the number of installations that are not triggered. Those wide-ranging or accidental can be useful in identifying malware and offensive campaigns that would otherwise go undiscovered \citep{stuxnetDiscovery}. 

\subsubsubsection{Complexity and Silo Breaking}
AI may help attackers make use of disparate data sources to identify especially vulnerable and lucrative targets but it is not clear how much they would improve over current approaches. A current approach is to spread the attack widely such as through spray-and-pray phishing or to have a worm attempt to exploit a vulnerability and either fail or succeed. It is also simple to conduct wide-reaching scans for vulnerabilities such as through Shodan \citep{shodanSummary}. Simple approaches such as these already help attackers identify vulnerable or lucrative targets across the internet.

\subsubsubsection{Access to Expertise}
If AI helps harden defenses, it could decrease the number of soft targets for attackers to choose from. At the status quo level, AI is helping human defenders harden code bases and make sense of alerts and suspicious activities in their networks \citep{googleBigSleep, securityCopilot}. At the reliable and independent level, defenders could further harden software and the entire digital ecosystem while providing low-resource organizations or even individuals with capable cyber defenders.

\subsubsection{Attackers Choose When to Strike}
Attackers choose to attack or not of their own volition. While that is always true, there may be pressures that influence their decisions to strike that can weaken the assertion that attackers choose when to strike. For example, a cyberattack may need to occur at a specific time in order to coordinate with other aspects of a campaign such as a ground invasion. In that case, the attacker has chosen when to attack even if the decision was not made by the cyber commander. As another example, a cyber team may be forced to the counteroffensive, such as requiring cyber attacks to repel an adversary’s ground invasion.

More often, a cyberattacker is free to choose the time of their attack. It may be when the defense seems to be weakest or when the impact would be greatest. Maybe a winter’s cold makes for a particularly painful time for the power grid to go down. Or maybe a new vulnerability is disclosed, making many targets weaker than usual.

\subsubsubsection{Speed}
AI agents may allow attackers to be more responsive to opportunities as they present themselves. Offensive development can already be damagingly fast as was seen when NotPetya and WannaCry repurposed the leaked EternalBlue vulnerability \citep{hern2017wannacry}. But it may be that increased speed does more to help defenders harden these targets of opportunity quickly than it does to help attackers. Attackers are already fast, so defenders may have more opportunity to improve from technological advancement, especially in the process of implementing defenses widely rather than in developing them \citep{lohn2022will}.

\subsubsection{Attackers Choose Their Goals}
Some argue that offense has simpler goals than defense \citep{slayton2017why}. Regardless of whether the goals are simpler, offense can choose their goal. Attackers could exfiltrate data, poison a model, flip a circuit breaker, deface a website, or perhaps wait to choose whichever is most convenient once they have entered a network. Defense needs to protect against all of those simultaneously.

\subsubsubsection{Availability}
The availability of reliable and independent AI decision-making could enhance this offensive advantage. If an offensive agent is granted the authority, it could potentially make these decisions quickly and without explicit command and control. It could take better advantage of fleeting opportunities. 

It is not clear how often opportunities are so fleeting that human decision making is really the limiting factor or how often command and control is difficult for an attacker. Some examples may be when targets are air-gapped industrial control system networks or classified networks. Without access to command and control, a reliable and independent attack agent may be able to decide on actions after accessing the target based on what it observes there in ways that would be challenging to automate.

\subsubsubsection{Access to Expertise}
Offensive agents may also have expertise about target systems that is hard to find among humans. There are many specialized systems such as industrial control systems, and organizations can be complex and opaque to outsiders. AI that helps attackers understand their targets, could improve the attacker's ability to identify impactful effects or actions. The same AI could also help defensive agents some, but defenders likely already understand their own systems and operations better than attackers do, so defenders would stand to benefit less.

\subsubsection{Targets are Easy to Find}
Targets usually need to be easy to find to offer their services. Becoming hard to find, as in a denial of service attack, can be the entire objective of the attack. But not all targets are easy to find. Some individuals or small organizations may try to be hard to find by disguising their identity or minimizing their digital footprint.

Where targets are typically easy to find, attackers typically strive to be undetected and unfindable. A possible exception are criminal attackers-for-hire who need to be findable to sell their services. Even in that case, the marketing of services may be easier to find than the services themselves and the attackers usually take many precautions to avoid being identified.

\subsubsubsection{Scale}
If AI increases the number of attacks, that would increase the number of opportunities for defenders to discover attackers. Defenders often use those discoveries to trace attacks back to their origin to identify attackers and their tools or infrastructure.

\subsubsubsection{Complexity and Silo Breaking}
The difficulty of finding attackers today provides an opportunity for defenders to balance the scales somewhat. If AI is able to help defenders identify and inspect suspicious behavior across the internet, it may do more to help defenders identify attackers than it does to help defenders whose targets are already easy to find.

\subsubsection{Defense Requires Reliability}
Defenders need their systems to remain operational and usable. Systems cannot be taken offline or restarted too often such as while installing security updates. And those updates cannot interfere too much with the usability or performance of the systems. For example, the patches cannot crash systems as happened in the 2024 Crowdstrike update that grounded airlines and derailed businesses across many sectors \citep{allyn2024what}. 

Offense is likely less impacted by poor reliability. To the extent that attackers only need to succeed once, they are more tolerant of failures. But those failures can lead to adverse impacts such as revealing their intent, infrastructure, or capabilities.

\subsubsubsection{Speed}
Uptime and reliability may place limits on how quickly and how often defense can adapt. If users only tolerate restarts once a day or if service level agreements limit the downtime of a network, then defenders may not be able to respond at the speed of attacks. Advances in live patching, where systems or processes can be patched without restarting, may be needed for defenders to take advantage of speed increases that AI promises to provide \citep{linuxLivePatching}.

On the other hand, defensive patching is often much slower than it needs to be and AI could potentially accelerate patch uptake \citep{lohn2022will}. Any AI advances that can help defenders be more confident in new fixes, or that can help defenders work through the process of implementing them, would have an outsized effect on hardening networks. Where the implementation delay comes from too few defenders to meet the long list of priorities, reliable and independent AI defenders could help speed defensive hardening \citep{jacobs2023enhancing}. Expert-level AI or greater may be needed for cases where the primary challenge is not simply implementing a long list of priorities. For example, significant expertise may be require to understand and assess the possible adverse impacts that updates could have across a network or organization.

\subsubsubsection{Improved Accuracy}
The reliability challenges that defenders face, such as in deciding whether an update will cause adverse impacts, could be reduced if AI helps produce more reliable code or services. That could happen if expert-level AI is more widely available. The opportunity for improvement may be limited though because many updates and defensive services are already developed by major organizations that employ teams of human experts. Superhuman AI may be required to substantially improve reliability of those updates.

\subsubsubsection{Scale}
AI may be useful for scanning, testing, and evaluating prospective updates or services as well. If AI can help defenders determine which changes are safe and which are not, then it could further reduce the disadvantages that defenders face from the demand for reliability. 

\subsubsection{Defense Has More Bureaucracy}
Attackers are often small organizations with a high degree of autonomy. That may be less true for nation-states, especially those in Western democracies that prioritize adherence to international law and norms of behavior. While other threat actors such as cybercriminals and terrorists also likely encounter some bureaucratic obstacles, they are likely less encumbered than their targets \citep{schneier2012how}.

Defensive bureaucracy can be substantial across government and industry. For example, acquisition procedures for cyber and AI tools can impose long delays, as can hiring decisions or approvals for defensive actions such as wiping a compromised device. Needless to say, an attacker can load their AI-enabled tool onto a classified network without the same bureaucratic process that defenders face to upload classified defensive tools. 

There is also tactical bureaucracy in defense. Approval may be needed from users or from leadership for decisions such as whether to kill a process, wipe an endpoint, or change the thresholds for spam or phishing filters.

\subsubsubsection{Speed}
Bureaucracy can be a rate-limiting step for defenders. When that is true, then speed of progress benefits attackers more than defenders. Bureaucratic processes can be designed to allow for rapid technological progress. Authority can be delegated for making tactical decisions. Acquisition approvals can be written to automatically approve new releases in an approved product line so that the most advanced technology is rapidly available to defenders. Bureaucratic reductions introduce technical risks and sociological or legal risks that may not always be acceptable, so bureaucratic delays may remain even with very advanced and fast AI systems. 

\subsubsubsection{Access to Expertise}
AI may also be able to provide expertise that is otherwise hard to come by. For example it is hard to find bureaucrats or acquisition officials who have the technical skills to rapidly evaluate new technologies. AI may be able to help resolve technical questions that these officials have. Where a technical expert may have been required to inspect new technologies to answer a bureaucrat's questions, AI may be able to assist instead.

\subsubsection{Defense Has Systematic Vulnerabilities}
Systematic vulnerabilities are those that come from a repeated method or approach. They can be different from systemic vulnerabilities, which are related to the overall system rather than just its components. A systematic vulnerability can be reproduced in many components without affecting an overall system, and a single systemic vulnerability can have a cascading effect impacting many components. 

Developers and defenders often use the same techniques or codebases, creating many instances of the same weakness across the cyber ecosystem. For example, if a vulnerable library is used within many projects then all of those projects could be vulnerable. Or if vulnerable code is provided from a common instruction website, then many developers might systematically reproduce those vulnerabilities in their projects and codebases \citep{fischer2017stack}.

\subsubsubsection{Bias}
Code weaknesses and defensive tendencies can be considered a form of bias. The vulnerability from a common instruction website may be learned just as easily by an AI agent as by a human, but if that AI agent writes a greater fraction of the code then it may reproduce that vulnerability more systematically than the human. In addition to developing secure or vulnerable code bases, the preference to choose one option for a defensive configuration or tactic from a set of comparable alternatives is a form of bias. When done at scale, that bias becomes a systematic decision, and if that decision is weak then it can be exploited systematically.

It may be possible for AI to reduce this risk by simply requesting that AI developers and defenders avoid repeating their designs and decisions exactly. Defenders could intentionally introduce variability into their designs and decisions to reduce the risks of systematic vulnerabilities in a way that is difficult to implement across human developers and defenders. AI developers may also reproduce code less exactly than human developers who often copy and paste code from repositories.

\subsubsubsection{Security}
AI systems themselves are systematically vulnerable in that they all tend to be manipulable in the same ways. Those pervasive weaknesses can be used to control their behavior. Those vulnerabilities affect both defensive agents as well as offensive agents, so it is not clear whether they advantage offense or defense in cyber engagements. But incorporation of vulnerable AI components in more systems probably increases the difficulty for defenders.

\subsubsection{Defense Has Systemic Vulnerabilities}
Apart from methodologically reproduced weaknesses, defenses need to protect complex and interrelated systems. Those systems often rely on key components that can be compromised. Large portions of the digital ecosystem may rely on a single library that can be vulnerable, such as in the infamous Log4j, Heartbleed, and SolarWinds vulnerabilities \citep{geer2003cyber}. Alternatively, a critical supplier may be vulnerable and a digital attack disrupting that supplier could cripple the complex chains that rely on it.

\subsubsubsection{Scale}
If AI increases the number of components and increases the complexity of organizations and systems, that could increase the potential for systemic vulnerabilities. The digital systems and the organizations built on those components or services would have more components or services that could be compromised.

\subsubsubsection{Complexity and Silo Breaking}
Part of the reason why these systemic vulnerabilities exist is that the systems and organizations are too complex for humans to adequately understand and evaluate. AI may be able to help assess complex systems such as by reviewing all software components or transactions across an organization's supply chain. If AI can help identify potential systemic vulnerabilities, that knowledge could help defenders mitigate them or at least better assess risk.

\section{The Character of Cyber}
\label{characterSection}
A forthcoming paper by Jason Healey, Robert Jervis, and Divyam Nandrajog tries to capture the set of characterstics that make the cyber domain what it is \citep{healeyCharacter}. They look across a variety of official policy documents and academic publications, recording notable pronouncements about cybersecurity or cyberwarfare. They also do not try to assess the validity of those pronouncements, simply compile them. Together, they are as complete a list of what is proclaimed to give cybersecurity its character as the authors of that paper or the authors of this one have seen. 

By assessing how AI may impact each of those pronouncements, we hope to start reasoning about how AI may change the character of cybersecurity and cyberwarfare.

\subsection{Tactically Moves At Network Speed}
Because of the speed of digital technologies, many individual cyber actions and activities already occur quickly. Individual tools or actions such as a network scan, a malware download, a file inspection, or permission change are unlikely to be made much faster by AI. But as Healey, Jervis, and Nandrajog note, tactical tasks in other domains can also be rapid. An artillery barrage might not be slower than a large file transfer or disk encryption. And some cyber tactics, such as finding a vulnerability, reverse engineering a binary, or writing an exploit or patch, can be more time-consuming than common tactics in other domains.

AI does not typically promise to make fast cyber tactics much faster. In fact, by digital standards, AI, with its large neural networks, is often considered slow. Perhaps AI at the expert level or above could redesign some cyber tools or algorithms to run more efficiently and improve tactical speeds in that way. But there are hard limits to how much some tasks could be accelerated.

For instance, information cannot be transmitted faster than the speed of light, and wires or memory devices in circuits charge at the RC time constant. Further, some algorithms are provably optimal and cannot be further improved. And attackers in particular might need to stay below speed thresholds to avoid detection.

AI does promise to make some slow tactics faster. Vulnerability discovery, reverse engineering, patch writing, code refactoring, and honeypot creation are all complex but tactical activities that AI promises to accelerate.

\subsection{Slower at Operational, Strategic Levels}
Cyber operations can be significantly slower than their high speed tactical elements would imply because of the need for humans and bureaucracies to navigate various potentially complex operational and strategic concerns.

That is likely to remain true as long as humans aspire to maintain control over the agents that do their bidding, be those agents human or machine. Some of those operational delays would probably shrink if humans delegate more decisionmaking to machines. But fully delegating decisionmaking to superhuman machines might not eliminate delays. The operational and strategic decisions in cyber are both technically and socially complex. They are challenging because human stakeholders may have diverse and unstated preferences or objectives that need to be sought out and considered as part of the decision-making process.

Operational and strategic pace is determined more by brazenness and willingness to delegate. In that sense, fear that an adversary might gain advantage by delegating operational or strategic decisions to their AI systems might drive an organization to delegate their own operational or strategic decisions to AI. Fear over adversary use of AI may do more to accelerate its use in operational and strategic decision-making than advances in actual capability.

\subsection{Cyberspace is a Scale-Free Network}
“Scale-free” is a technical term referring to a type of network where a few nodes have many connections but most are only connected to only a few \citep{barabasiScaleFree}. Those networks are resilient in the sense that impacts on most nodes do not drastically affect the whole network. But there are a relatively small number of single-point-of-failure nodes that have outsized impacts on the entire network.

AI could upend the structure of the internet and digital networks in various ways. If AI becomes pervasive and it is provided by an oligopoly or monopoly, then the internet and digital technology more generally might become even more centralized than it is today. More traffic could flow to and from one or a few organizations (i.e. model developers) with less need for smaller or medium nodes to interact among themselves because the information they need is contained in the models. Rather than navigating to Wikipedia, the weather network, news sites, and government pages directly, a user may request that information from an AI services that then reaches out to those pages on their behalf. Rather than a “scale-free network,” this configuration would appear more as a “hub-and-spoke network.”

An alternative possibility is that AI becomes pervasive but that it proliferates widely and runs efficiently. Open-source and low-resource models and agents that are reliable and independent could be built directly into systems, reducing or removing their need for network access. Most of the information that they would need to operate or that their users require of them could be self-contained.

The future internet could be some combination of these two scenarios. An example presaging this combined future is Apple’s agreement with OpenAI \citep{kerner2024openai}. Apple’s iPhones have a small efficient model that does as much as possible on the phone without network access. It only sends more difficult tasks to the more capable models at OpenAI. Some information that the iPhone previously might have needed to retrieve from the internet is now contained within the device’s internal AI model. And additional information that is needed for more challenging issues may be requested from OpenAI’s servers as part of its process for responding rather than in a request directly from the iPhone.

These changes could upend much of the interdependence of today’s internet where nearly all targets are connected and reachable at all times. Proliferation of models or agents that are efficient, reliable, and independent could lead to many systems being removed from networks that they no longer need, or reducing the number of network connections they need to function. Those reductions could convert those systems into harder targets. 

\subsection{Cyberspace is Human-Made and Adaptable}
In contrast to other domains such as land, sea, air, and space, cyberspace was made by humans and is comparably easy to change.

With the amount of code that AI has already contributed to, one can argue that cyberspace is already not solely “human-made,” but the point is not so much about whether humans or machines make cyberspace. The point is that it is made to reflect human preferences and that it can be changed to better reflect those preferences.

If AI were to develop its own preferences and design digital systems in accordance with those preferences, then this point would need to be revisited. Even limit-breaking AI systems do not need to have preferences of their own and would presumably execute tasks according to guidance from their human operators and designers. Without sentience, the spirit of cyberspace being human-made does not change simply because the mechanics of it are conducted by AI.

What does change is the adaptability of cyberspace. If AI can enable code and configurations to be changed at a faster rate, then cyberspace would become more adaptable. This is already happening somewhat at status quo levels of AI and is expected to accelerate as AI progresses further. 

At some point, adaptability might interfere with usability. Just as human programmers can be frustrated when programming libraries or services are updated too quickly, AI programmers would struggle to build systems if the components in them or services they are built on change too rapidly. That effect could further push human developers out if they are unable to keep pace with changes in libraries and services. Additionally, at the user level rather than the developer level, if updates and changes require downtime or changes to user behavior, then too many changes could interfere with usability or service level agreements.

\subsection{Cyberspace is Unfathomably Complex}
The amount of code, connections, and possible configurations within cyberspace make it impossible for any human to understand completely. Humans do understand much of it through various levels of abstraction. They understand that digital devices are made of silicon that is patterned into circuits that are controlled by firmware that receives signals from applications that are connected to other applications through the internet and its collection of service providers and their agreed-upon protocols. People understand how these pieces fit together to make cyberspace, and individual humans can understand the detailed intricacies of a few of these but not all at the same time. 

AI specializes in managing complexity. It can take in large amounts of information and find patterns or pick out salient elements. It is already helping humans make sense of aspects of this complexity such as by analyzing and explaining code bases and digital binaries or by observing and explaining network traffic patterns. 

Further advances may help make even better sense of cyberspace’s unfathomable complexity although the volume of data and information across cyberspace make it hard to imagine even AI making sense of large segments of it. It is also hard to imagine that level of detailed analysis being necessary. The abstracted understanding that humans have is sufficient for many purposes. But perhaps a more advanced intelligence would find a more detailed understanding of cyberspace’s complexity valuable in ways that are hard to envision today.

At the same time, the prospect of AI writing vast amounts of new code could substantially increase the complexity of cyberspace. If that code is written by superhuman levels of AI, or perhaps even less capable AI, some of that code may not be easy to interpret for humans. The code or systems could be complex in design or simply written in difficult to interpret languages. AI could also conceivably contribute to standards or protocols, adding complexity beyond to the structure of the digital ecosystem, not just its components.

AI, even at the status quo level will help make sense of cyberspace’s complexity, but even limit-breaking AI may not relieve its incomprehensibility to an impactful degree. Cyberspace will likely remain understandable and understood in the abstract and in its individual components but unfathomable in its entirety. 

\subsection{Cyberspace Contains Inherent Vulnerabilities}
Vulnerabilities are pervasive in cyberspace. Although they are probably not infinite, the rate at which humans discover vulnerabilities is consistent with an infinite supply of vulnerabilities to discover \citep{lohn2022will}. 

AI could reduce the prevalence of vulnerabilities in a few ways. For one, it could be used to review and refactor large code bases. That is happening to some degree at the status quo level although AI may currently be increasing the number of vulnerabilites rather than reducing them \citep{rabbi2024AIvulns}. At the reliable and independent level, AI agents could be trusted to do basic cleanup across vast code bases to implement basic security practices \citep{darpaAICC}.

A more advanced task is vulnerability discovery. Status quo level AI is already helping accelerate human vulnerability discoverers but expert or superhuman AI might find vulnerabilities that humans would not otherwise find \citep{googleBigSleep}. But that is only a benefit to cybersecurity if the tools do not help attackers find vulnerabilities that defenders cannot patch for some reason. Those reasons could include that there are too many vulnerabilities for defenders to find all of them or that patching is difficult, impossible, or is undesirable such as if it impedes performance such as can be the case for hardware vulnerabilities \citep{lohn2018meltdown}.

Defenders may also benefit from AI systems that can design software that is more secure from the start, possibly even software that is provably secure. Writing code in memory-safe languages is one near-term approach to hardening the digital ecosystem \citep{rohlf2023memory}. Significantly advancing provably secure design might require superhuman levels of AI, but might be possible with large numbers of expert level AI software designers \citep{woodcock2009formal}.

Reliable and independent AI defenders may be able to harden large volumes of code, performing tasks that are not especially difficult but that are too numerous for humans or that humans find boring or otherwise unappealing. That could eliminate many known vulnerabilities from the digital ecosystem. Further advances toward the expert levels may be needed to discover many as-yet undiscovered vulnerabilities. Discovering vulnerabilities helps defenders harden their products, but if AI makes undiscovered vulnerabilities plentiful, then attackers probably benefit more than defenders \citep{lohn2022sword}. Further advances toward superhuman AI may tilt the scales back to defenders if it can make software that is provably secure, making those undiscovered vulnerabilities hard to find again.

\subsection{Low-Impact Reversible Effects of Capabilities}
Cyber attacks have typically fallen more in the category of nuisance than disaster, but that may be becoming less true. Some cyberattacks have resulted in disaster-level financial impacts, such as the billions in losses from NotPetya \citep{mcquade2018untold}. Attacks on critical infrastructure such as the Colonial Pipeline or across the healthcare industry have also grown more common \citep{easterly2023attack,riggi2020ransomware}. 

It is also unwise to conclude that cyber attacks will continue to have low impacts based on the precedent that, during a period of relative peace between cyber powers, there have been few high-impact strikes. But limited impacts may not be only due to limited intent, defenders may also be reluctant to overly digitize or connect especially risky systems such as some military equipment or aspects of critical infrastructure. If advances in AI make digital systems more vulnerable, then defenders can choose not to expose their systems to digital attacks. But if humans grow to rely more on AI-enabled systems, then resilience during and following cyberattacks could decrease. 

If the historically low-impact effects of cyber is because the most capable threat actors have been somewhat restrained, then advances in AI could change the degree of impact to expect from cyber attacks. Where even rogue nation-states and cyber criminals have been somewhat reluctant, cyberterrorists may not be. Terrorists have historically been too low-skilled to execute damaging attacks \citep{rid2012weapons}, but that could change if AI attack tools are capable enough and proliferate widely enough. But proliferation of attack tools may not be so useful to terrorists if defensive tools allow defenders to effectively harden their systems and the digital ecosystem more broadly.

\subsection{Attack is Lesser Included Case of Espionage}
A cyber campaign that is intended to cause damaging effects is difficult to distinguish from a campaign that is considered legitimate espionage. In a sense, it is the defender’s duty to repel both legitimate espionage and illicit effects. So while true, and perhaps important at a strategic or diplomatic level, this distinction, or lack thereof, may be less important at a technical level.

If AI shortens the number of hours or days for a cyber campaign, then prepositioning destructive offensive capabilities under the guise of espionage would be less necessary. Attackers could wait to gain access to conduct their attacks only when needed. At the same time, the benefits of persistent espionage would likely remain or increase in if AI advances further, so espionage campaigns will probably remain popular at any level of AI advancement. And those espionage campaigns will mostly continue to be easy to convert into illicit effects campaigns.

In some cases though, AI may be easier to infer intent from a capture AI implant than from current cyber tools. If defenders can query the intruding agent or evaluate it in a safe sandbox, they may be able to determine whether or when it would conduct illicit activities rather than legitimate ones. But confirmation that a cyber agent is legitimately meant for espionage rather than destruction is only relieving if it would be difficult for attackers to update or exchange the agent to make it more aggressive. In some cases, the agents might not be able to receive those updates so intent could be inferred, but the baseline assumption going forward should be that those updates would be trivial.

\subsection{Fast Pace of Technological Change}
Digital technologies improve and change rapidly, as do the techniques that attackers and defenders use. 

AI is already increasing the rate of innovation in terms of humans’ ability to design and write new code and components as well as the types of approaches that attackers and defenders use. That acceleration should be expected to continue but perhaps not ad infinitum because there could be performance and interoperability challenges that arise from changing too many interrelated components and services too rapidly. A service can break when an underlying service is changed or updated. And the need for stability across the global ecosystem may exert a pressure against rapid change among its elements.

\subsection{Universal Interconnection and Dependence}
A global internet and a unified digital ecosystem built on a complex digital supply chain make people, organizations, and nations reliant on one another in ways that can be difficult to assess and understand. That may be reducing somewhat as the global internet becomes more segmented with countries decoupling from one another such as through export controls or censorship and as companies try to reduce their exposure to supply chain risks.

If AI reshapes the internet to more of a hub and spoke network rather than a scale-free network, then the network-level interconnectivity could further decrease. That could also happen if reliable and independent agents disconnect from networks. 

On the other hand, if AI creates more components, drawing on a greater number of more diverse resources, the digital ecosystem could become more interconnected and interdependent. Software could incorporate libraries or coding practices that a human developer would not be aware of because it was originally developed for obscure purposes or commented in different languages.

\subsection{Permissionless Innovation and Connection}
The internet and many aspects of the digital environment were designed to be easily accessible and easily contributed to. This is apparent in the freedom of exploring the internet and in the ubiquity of open source contributions.

There is a suite of technologies and procedures such as robots.txt or CAPTCHA that have helped to control the roles of non-human digital agents to some degree. Some of these techniques are growing less effective against even status quo AI. It may be possible or necessary to develop more robust proof of humanity \citep{seldmeir2021DigitalID} for some applications.

As AI grows in prominence, the battleground for openness may be getting data included in training sets or the distribution of models, systems, and services. Some developers or nations will choose to include or exclude various data sources. They may also use more aggressive data cleaning procedures, including censorship. A Chinese model will not treat all international issues the way an American model will. And export controls or know-your-customer policies could also limit which actors and nations have access to which capabilities.

\subsection{Fuzzy Borders}
Digital signals are not so easily constrained to geographic boundaries. Certainly countries and companies can draw boundaries that are either digital such as through firewalls or geographic according to physical infrastructure. For that reason, cyber is not borderless, but borders are fuzzy.

If AI accelerates cyber offensive and defensive engagements to machine speeds, then the time to send signals between distant countries could become significant. Communication delays have long been a limiting factor in high-speed trading for example and may already be significant for race condition vulnerabilities for example \citep{fischer1992race}. At high speeds, the borders would remain fuzzy but geography could play more of a role than it currently does.

In the nearer term, AI has already become a diplomatic battleground. Some geographical regions may be restricted from accessing or utilizing the most capable systems. This scenario is more likely if access to the most capable AI is limited to a few organizations rather than if it proliferates widely.  

\subsection{Tied to the Physical World}
Digital technologies and data may seem ephemeral but they exist within physical components such as circuits, fiber optic cables, and network routers.

That is true for all digital systems whether they are AI or not, but the significance of the digital components depends on the technology. Currently, the most capable AI systems rely on expensive computer chips that are dominated by a single provider. Those chips are built using transistors that are fabricated by another dominant provider in fabs that require components designed and constructed by still other companies. And the most advanced models are trained and operated in physical datacenters that are large in size and have substantial power and cooling demands.

The physical aspects of AI have been a more significant aspect of cyberspace as related to AI than for most of previous digital technologies. That could continue or grow if AI continues to be extremely resource intensive. If instead, sufficiently capable AI becomes less resource-intensive, then that physical dependence would fall toward levels consistent with prior digital technologies. 

\subsection{Ease of Copying Information (and Capabilities)}
Digital technologies, at least software and information if not digital hardware, can be rapidly reproduced.
Copying information is already very efficient, so AI is unlikely to make it much easier. But to the extent that AI is encoding and aggregating data and information that would otherwise be contained in humans or distributed across many disparate sources, AI may make the ease of copying digital information more damaging. 

For example, imagine an AI model trained on a large corpus of classified information. Today it is difficult to compile information that is kept in separate compartments with various permission restrictions. If that information is condensed in a future AI model, copying that single model would be similar to stealing everything it was trained on.

As another example, imagine a personal AI assistant that observes all of a person’s interactions and behaviors to learn their preferences and interests to behave accurately on their behalf. The existence of that model would allow for attackers to easily copy information that is not collected and retained today. This is already happening as phones and wearables collect data but could be further exacerbated either by AI or to better enable AI. 

\subsection{Dominated by Private Sector}
Much of the infrastructure and expertise in the digital world is housed in the private sector. A possible exception to this is that governments, which are the only organizations with legitimate offensive authority, have the most capable offensive cyber teams, tools, and infrastructure.

AI systems are similarly dominated by the private sector. Private companies have the budgets and expertise to develop AI systems that governments have not been able to match. Governments are likely to be able to exert some influence over those companies and to cooperate with them. Governments would also be able to provide the companies useful information such as threat intelligence while also receiving information such as tips about suspicious accounts. Governments may also be able to provide more or less permissive regulations or even shut down companies if necessary.

Even if governments are unable to produce or acquire cutting-edge AI systems, they may may be able to leverage their expertise, tooling, and authorities related to offensive cyber to maintain their dominance over the private sector on the offensive side of cybersecurity.

\subsection{Overall Attacker Advantage}
A common maxim in cybersecurity is that attackers have the advantage. As understanding AI's impact on this maxim is the objective of this entire paper, we refer the reader to the discussions in every other section.

\subsection{Difficulty of Quick or Exact Attribution}
Digital identities and actions are separate from those of the actual perpetrators, making attribution often difficult or slow.

The more that the human operator is removed from the process of a cyber attack, the fewer clues there are to tie that human to the attack. If the same AI is used by Russians and Chinese to write an attack script, then that script will be less useful in distinguishing between the two. And AI could be directed to create scripts or conduct attacks how other threat actors would for false flag operations.

From the defensive standpoint though, AI may be able to look across more data from more complex sources to compile clues and make a case for attributing the attack to one perpetrator over another. Many clues would remain even if the entire attack was conducted by an AI, such as motives, location and ownership of attack infrastructure, and communications of leaders or commanders. 

If AI manages to accelerate the process of attribution, it could be concluded on more useful timelines rather than waiting months or years, during which public outrage may have softened and legal or technical retribution may face additional challenges such as extradition. 

\subsection{Hard to Directly Observe}
Digital information and digital effects can be difficult to observe. Intruders can be stealthy, attack code does not show up in overhead satellite imagery, and even changes to the settings of a centrifuge may go unnoticed \citep{zetter2014unprecedented}.

\begin{sloppypar}
  AI has long been used to scan systems and files for potentially malicious code or activities. Presumably AI will continue to improve, although historically it has been challenging for defensive AI to keep pace with evolving offensive tactics \citep{patchPrioritization}. It is not clear whether more advanced AI would make the already widespread use of autonomous scans more able to identify malicious files and actions or if advanced AI would make those files and actions harder to detect such as by behaving more like genuine users or by writing polymorphic malware.  
\end{sloppypar}

Observing attacker activity also depends somewhat on the degree of control over access to AI for cyber. If attackers need to submit their requests to be conducted on an AI company's servers then those companies could monitor and possibly restrict threat actors prior to or during attacks and could log activities for forensic investigation. Even if those services are provided by more permissive companies or rival nations, they may all agree to suppress some types of attacks or threat actors such as terrorists. 

If capable AI is widely distributed and easily accessible, then observing or interfering would be more challenging. Limitations or guardrails built into models are not currently promising for restricting malicious users. Further, the capabilities that might be needed for defense such as vulnerability discovery and penetration testing are similar to the capabilities needed for offensive use.

\subsection{Capabilities Are Transitory and Have Hard to Predict Effects}
Offensive tools and techniques can often be rendered ineffective with one or a few simple defensive changes. As a result, offensive capabilities may not be useful for long. That can put pressure on the offense to use their capabilities because they might not be able to use them in the future. 

That ability to quickly render offensive tools and techniques ineffective, along with the complexity and interdependence of cyberspace also makes it difficult to predict offensive effects. A campaign or attack may fail unexpectedly. The attack may cause more or less damage than desired to a target, or it may cause collateral damage to unintended targets.

If AI finds more new vulnerabilities quickly and accelerates the patching process, then offensive capabilities may become even more transitory. Even simply accelerating normal updates or changes to target systems can unintentionally disarm offenses \citep{ablon2017zero}. At the same time, if it becomes easier for offense to find new vulnerabilities and create new tools and techniques, then offense would find the transitory nature of its tools and techniques less problematic. Attackers could expect to create capabilities at the times when they are needed rather than needing to maintain a repository of prepared options. 

Regarding uncertainty of effects, attackers may be able to reduce uncertainty by conducting more attacks. If each attack is uncertain to produce the intended effect but more attacks can be attempted in the same number of hours or weeks, then attackers can be more confident of at least one success in those days or weeks.

But that increased number of attacks could also lead to an increased probability of accidentally causing too much effect that might be more escalatory or damaging than intended. It also increases the probability of collateral damage. On the other hand, an increased number of attacks could also increase the opportunity for defenders to identify and disrupt attacker operations.

\subsection{Adversary Forces in Constant Contact with Few if Any Operational Pauses}
Attackers and defenders, especially large ones such as nation states, large companies, and criminals, are constantly launching or repelling cyber attacks. That is unlike most kinetic conflicts in other domains where engagements are rare.

Some of that constant contact is to preposition to be able to quickly cause damaging or destructive effects, which would be less necessary if attackers believed that AI could help them initiate and conclude attacks quickly and reliably. But prepositioning for destructive effects does not motivate most cyber engagements. The desire to continuously collect information for intelligence purposes would probably not decrease with advanced AI. Cyber criminals would also probably continue to be motivated to conduct regular attacks.

If AI increases the number of attackers or the number of agents per attacker however, the number of contacts could increase. What feels like constant contact today may not look like constant contact in an age of widely available cyber attacker agents that act reliably and independently. Where today there are many botnets that mostly perform simple automated messaging as in a DDoS attack, in the future, those bots may each be capable of taking a creative course of action that defenders would need to account for individually. 

\subsection{Immediate Intercontinental Proximity to National Sources of Power}
Digital signals can travel around the world at the speed of light, reaching their targets much more quickly than ground troops or missiles can. Additionally, there is often little difference in the digital world between reaching a nation’s border and its capital or its industrial centers.

If digital engagements were to accelerate sufficiently, and the speed of decision making becomes a decisive factor, then the speed of light to move information between geographic locations may not be fast enough. Those time delays could be much longer than the speed of light if that signal needs to pass through an obfuscation network to disguise its destination and if the contents need to be assessed and filtered at firewalls such as to prevent counterattacks.

Additionally, if AI reshapes the internet, it could make some important targets less digitally reachable. Targets could use their own reliable and independent AI systems or agents and disconnect from the internet, or reduce their number of connections, making them more difficult for attackers to reach. 

\subsection{Advantage Comes From Use of Capabilities Not Possession}
Cyber offenses differ from nuclear missiles for example, where the threat of use may be more effective than their actual use. Providing too much information about a cyber threat allows defenders to render it ineffective. 

That perspective is somewhat overstated given that nations and companies do account for the capability of adversary cyber teams despite not knowing specific adversary cyber capabilities. Defenders presume that a sufficiently advanced hacking team can break into any digital system with enough time and effort. And there are estimates for how many of those teams are available to each country, as well as estimates for how capable and sophisticated various criminal groups are. 

With advanced AI systems, it may be even easier to accurately gauge the ability of adversary offenses. If AI attack tools become expert level or above, then penetration testers could use AI tools that are the same or similar to the tools and expertise that the adversary has available. Those tests may not exactly represent adversary capability if the attacker has additional data, tools, or expertise that significantly contributes to their capability. The adversary may also lack important data or tools. And if access to the most capable AI systems is controllable, then it may be denied to attackers or adversaries. 

If the results of the red-teaming do not definitely say whether a target is hardened or vulnerable, then advantage may still come from use of capabilities rather than their possession even if AI advances significantly.

\subsection{Adversaries Routinely Use Capabilities Mostly Below Level of Armed Conflict}
Offensive cyber has become a preferred tool for causing limited effects that hamper its victims but not to a degree that leads to escalation. Cyber attacks have gained a reputation for being a penalty-free form of aggression.

AI is unlikely to significantly change the value of digital espionage or the financial benefits to criminals, so cyber will likely remain in constant use at low levels of impact. Aggressors will continue to use cyber attacks to achieve as much as possible without excessive punishment.

AI could increase the risk of crossing the threshold into armed conflict though. More frequent attacks from a larger number of attackers, potentially with reduced control and oversight, increases the risk of accidental escalation. AI may also be poor at understanding the responses of its human victims, who may be culturally different from those in its training set. But even if AI better predicts the human response so that each individual attack becomes less likely to cause escalation, the risk of escalation could still increase if the volume of attacks increases, especially if the offensive agents are prone to errors. 

\subsection{Low Barriers of Entry}
Offensive cyber tools are already available to, and usable by, individuals with very little technical knowledge or skills such as hacktivists, thrill seekers, insider threats or low-level criminal organizations. More advanced persistent threats require skills, knowledge, teams, and infrastructure, but even that is typically a small investment as a fraction of national or military budgets.

AI promises to disseminate expertise in ways that could make low-skill attackers even more capable. AI could provide them with the expertise to conduct more sophisticated attacks. At the same time, AI promises to harden code bases and prospective victims, potentially rendering those attack capabilities insufficient.

A possible progression is that proliferation of reliable and independent AI defenders harden their systems enough that reliable and independent AI attackers struggle. With further advances, proliferation of expert AI attackers may advantage attackers against expert AI defenders. But superhuman defenders may help construct defenses that are difficult for even superhuman attackers to overcome. 

\subsection{Superiority Is Fleeting}
The digital world’s speed of change, its reconfigurability, and its low barriers to entry imply that any cyber advantages one has may not last for long.

AI is likely to further accelerate the speed of change and reconfigurability of the digital world and possibly further lower the barriers of entry for cyber actors. Additionally, AI technology itself is rapidly changing and may continue to change rapidly. A country or company that has the most advanced AI at one time may only a few innovations away from having their technology rendered obsolete as other countries or companies continue to innovate.

AI is also a mechanism for rapidly disseminating expertise. A nation or organization that is advanced today because of its teams of many exceptional experts would not be so exceptional if other organizations gain access to AI with comparable expertise. AI could level the playing field on both offensive and defensive sides of cybersecurity.

\subsection{Capabilities Are Substantially Cheaper than in Other Domains}
The cost to develop cyber capabilities is typically low compared to planes or tanks. The cost is even lower to produce many copies of those capabilities once they have been developed.

AI may be expensive to produce or to use, but it is likely to further reduce the cost of cyber capabilities. The costs of developing AI is distributed across nearly all industries so cyber applications can benefit from economies of scale offering relatively low cost AI systems. But it is possible that highly capable cyber systems would require generation of huge data sets that could be expensive to develop or to train on \citep{lohn2023autonomous}. In that case, cyber capabilities could conceivably become more expensive than they are today.

\subsection{Capabilities Can Be Rapidly Regenerated}
When defenders repel or thwart a cyber attack, it is often only a temporary victory because cyber capabilities can be quickly reestablished. As examples, it takes comparatively more time to rebuild a submarine, and it takes more time to reestablish an industrial base after a tank manufacturing depot is bombed. On the other hand, when a defender repels a cyber attack, they may do so in a way that provides a lasting improvement to their defensive posture.

If AI accelerates the development of cyber capabilities, then that regeneration will occur even faster than it does today. The regeneration times may become a more significant fraction of the attack campaign delays though if there are bottlenecks that are more difficult for AI to accelerate. For example, developing infrastructure for attack can require creating accounts and identities could be slow to establish, especially if they require the legitimacy of human followers. And to the extent that infrastructure providers are a component of the broader cyber defensive, they have some ability to control these reconstitution timelines through know-your-customer policies for instance.

\subsection{Attacks Might Lead to Catastrophic Effects}
Although cyber attacks have a reputation for achieving small temporary effects that stay below the thresholds for escalation, they have the potential for more significant effects. In principle, they could be used to poison water supplies, destroy power generators, or initiate kinetic strikes. A collection of small attacks can also lead to catastrophic effects such as gradual democratic backsliding.

An increase in the volume of attacks could increase the likelihood of accidental catastrophes. Additionally, the availability of sophisticated attack tools and techniques to cyber terrorists could increase the risk of intentional catastrophic cyber attacks. Presumably, upskilling cyber terrorists would occur alongside upskilling cyber defenders. The defenders who stand to benefit most are the expertise and budget constrained small and medium sized organizations that include much of critical infrastructure. 

Integration of AI into more critical systems also increases the potential for catastrophic effects. Today, human operators provide a non-digital barrier between attackers and effects. Humans also provide a means of resiliency during and following attacks. AI is notoriously vulnerable to attackers, so integrating more AI components may increase vulnerability to attacks while potentially decreasing resilience to them.

\subsection{Tactical Success Tied to Agility and Initiative}
The ability to change course and adapt to opportunities as they present themselves, and the ability to adjust to attacker and defender behaviors or tactics, are important to succeeding in individual attacks and campaigns.

AI at the reliable and independent level or above could be authorized to make tactical decisions in real time without additional guidance. That can be true for both offense and defense. It may lead to mistakes simply as the number of attacks and defenses grow. That would be especially true if decisions are delegated to the AI because of speed and scale rather than because of accuracy or reliability. The penalty for mistakes may be higher for defenders who need to avoid adversely impacting usability and uptime for stakeholders. Attackers mainly need to avoid being discovered and avoid unintended escalation or collateral damage.

\subsection{Strategic Success Possibly More Tied to Audacity and Initiative}
Strategic benefits in cybersecurity have tended to accrue to the more aggressive actor. That may be because cyber threat actors are still toeing the line and have yet to discover where their aggressions will invite costly retaliation. While attacks continue to result in small penalties, more audacious attacks result in more strategic benefit.

Perhaps with a greater number of attacks, it will become clear what severity or frequency of attack will result in costly retaliation. It is possible that an increased number of attacks, none of which more audacious than typical attacks of today, could drive a nation to respond militarily. The resulting destruction could easily make those cyber attacks a strategic misstep. If the cyber red lines become clear then aggressors may choose more audacious attacks but it would be less true that further audacity and initiative would lead to strategic success. But AI might not clarify red lines, and audacious AI attacks might continue to reap strategic rewards.

It is yet to be seen how the public or nations will respond to AI-driven attacks. An AI agent may be more restrained and controlled than a traditional bot or worm. But people may recoil at the perceived increased threat of an AI-directed attack rather than a human-initiated one. Its artificiality, or for any of a variety of other reasons, could make audacity less beneficial for AI attacks. 

\subsection{Difficult to Deter}
Some claim that cyber attacks are difficult to deter for reasons such as that the difficulty of attribution allows attackers to claim innocence after an attack and that secrecy around cyber capabilities keeps deterrers from knowing about the capabilities they would need to deter.

If AI is being used to make decisions then it may be easier for deterrers to understand its decision calculus than it would be to understand human attackers. If defenders have a similar AI system or have captured the actual attacking agent, they could inspect it to determine what its responses would be to their defenses or provocations. But if the offensive agent or its guidance can be updated by the attacker, then those defensive investigations may be of little value.

Rather than understanding the response, defenders could use their own attacking agents to better understand their vulnerability. Red teams today try to imitate adversary abilities, but the human expertise to imitate expert adversaries is limited, so few organizations can be red-teamed appropriately. With widely available AI attackers, many organizations could have a better understanding of the adversary capabilities they need to deter.

Another reason for difficulty in deterring cyber attacks comes from attackers presuming that they can plausibly claim innocence following an attack. It is unclear how AI may affect cyber attribution. AI’s ability to ingest large amounts of complex information may enhance attribution. But if many attackers use similar AI tools for attack, or if AI is able to conduct attacks in the style of another attacker, then attribution may be more difficult in the future. 

\subsection{Defensive Success Does not Discourage Attackers}
Attackers only need a single success, so attackers can continue attacking until they succeed. A counterargument is that defensive successes may not simply thwart an attack, it may also allow defenders to disrupt attacker infrastructure or create alerts that the entire defensive community can share to identify the attackers and block their activities.

Human attackers also do get discouraged by repeated failure. Reliable and independent AI attackers on the other hand truly would not be discouraged. If defenses are imperfect, a large volume of attackers working 24/7 at a high rate of reengagement without tire or frustration will eventually find those defensive imperfections. But those repeated failures would still pose risk to the attackers that defenders could disrupt their infrastructure or share successful defensive tactics, techniques, and procedures with the broader defensive community.

\subsection{Difficult to Warn of Attacks}
Cyber attacks can be spontaneous and it can be difficult to observe preparations. If malware is developed on adversary systems and analysis occurs only on their networks, then there may be no signal of impending attack before it begins. That is part of the reason why nations defend forward and persistently engage; they feel the need to observe adversary preparations and plans before attacks are initiated. But some warning is possible when preparations happen in target networks such as from pre-placement of malware and attack tools or during network and target reconnaissance.

If AI shrinks the timelines for intrusions, tool development, or attack plans, then warning timelines could further decrease even if indicators remain detectable. Today, levels of alert are often tied to levels of political tension or escalating events more so than to technical indicators of attack. Those non-technical indicators will likely remain available to guide levels of alert. These political activities and escalating events from mainly non-digital domains are not as likely to be accelerated as much by AI because of their reliance on social or physical timelines such as public statements or the positioning of troops.

\subsection{Signaling Intent is Problematic}
It can be difficult to interpret the purpose of an offensive cyber tool or intrusion. An intrusion or malware may be meant for legitimate espionage purposes or it may be intended for illegitimate destruction. Further complicating matters, the attacker may not have a decided purpose, preferring to leave both options open. An intrusion into adversary networks may also be to defend forward or to cripple threatening capabilities, making the distinction between offense and defense unclear. And an individual or organization may penetrate a company to improve its defenses and collect a bug bounty.

If decision-making is delegated to AI, and that AI can be captured and interrogated, then it may be possible to determine its intent better than cyber implants today. That provides some hope that AI could resolve some of the intent problems of cyber conflict. But in cases where the AI can be rapidly updated, as is probably often the case, the actual strategic intent of the human attackers will remain as unknown as it is currently. These interrogation techniques may be more useful for retroactively discovering intent forensically than preemptively assessing implants.

\subsection{Likely to Be First Strike Weapons}
That cyber capabilities can be quickly defeated by a warned defender, and because cyber attacks are often less escalatory, some expect that they are preferentially first strike weapons \citep{healey2020escalation}. Incentives to employ cyber attacks in crisis could increase the risk of escalating to conflict, but cyber attacks may also act as a pressure release valve helping to de-escalate.

If AI allows attackers to expect to achieve access and effects on a shortened timeline, then those attackers will have less “use it or lose it” pressure. They will have less incentive to conduct cyber preparations and pre-placements that could potentially be escalatory during crisis.

But cyber attacks are likely to remain attractive options for low-stakes, short of war, actions. There are many levels of impact or destruction that are possible and cyber attacks have come to be seen as relatively low-stakes compared to kinetic alternatives.

\subsection{Surprise Is More Important}
Because defenders can typically thwart offense if provided thorough enough warning, attackers rely on surprise. That is not always true as evidenced by attackers who prefer to engage persistently rather than rely on surprise. Additionally, having previously conducted attacks, such as Russia shutting off Ukrainian power twice, does not give defenders certainty that they can prevent similar attacks in the future.

If AI enables offense to dominate tactically, then attackers would not worry about announcing their targets and effects and would benefit less from surprise. Defenders would believe that attackers could conduct the attacks they threaten. Short of disengaging those targets or their digital components, defenders would be unable to defend those targets. That disengagement may be sufficient to achieve the attacker’s operational and strategic objectives without actually conducting a cyber attack. 

But if attackers do not develop so much tactical advantage, then the element of surprise will still provide attackers significant benefit. It is far from clear that AI would provide such significant tactical advantage to attackers.

\subsection{Easy for Nations to Leverage Proxies}
Nations route their attacks through other countries and organizations to avoid detection and attribution. They also allow criminals and individuals to conduct attacks that align with their objectives and interests. Those criminals and individuals may even be employed by nation-states during the day or be instructed and tasked by the nation-states.

Reliable and independent AI attackers could be created and tasked by nation states or criminals and placed throughout the world. These agents could have few observable characteristics that tie them to their designers and deployers, making them scalable alternatives to existing proxies.

Attackers may also be able to co-opt the attack agents of others or others’ well-meaning software or agents. AI systems are notoriously vulnerable to poisoning and subversion. For example, attackers could place a prompt injection attack in a common data source driving benign programming bots to create privilege escalation pathways in their code or driving trading bots to pump specific stocks. More integration of AI could increase the opportunity for unwitting proxies. 

\subsection{Offense and Defense Similar, Inform One Another}
Many of the tools and techniques needed for offense and defense are the same \citep{buchanan2017cybersecurity}. For example, discovering vulnerabilities is critical for both offense and for defense. Penetration is a key objective of offense and a frequent testing technique for defense. And especially as defenders defend forward by penetrating the networks of attackers and disrupting their attack capability, cyber defense has begun to look even more like cyber offense. Some argue that the similarity between offense and defense invalidates the premise of an offense-defense balance \citep{valeriano2022failure}.

It is easy to overstate the similarity between offense and defense. Rewriting code to remove vulnerabilities or converting it to memory-safe languages is not an offensive act. Nor is instituting two-factor authentication or restricting permissions to files, folders, or accounts. And wiping the master boot record, weakening an encryption mechanism, developing deniable infrastructure, or introducing vulnerabilities to code are not defensive acts.

To a large degree, the activities that are similar between offense and defense do not depend on whether they are performed by a human or an AI, so AI will not significantly change how similar offense is to defense. What may change is how prevalent and important the actions are. If vulnerability discovery, penetration testing, and defending forward become more central to cybersecurity, then offense and defense will appear more similar. If cybersecurity becomes centered on a race between defensive agents trying to harden code bases and configurations against offensive agents trying to introduce vulnerabilities and weak configurations, then offense and defense will appear less similar.

\subsection{Conceptual Confusion and Lack of Precise Definitions}
People often struggle to understand and communicate the elements of cyberspace and the activities within it. For example, experts debate what constitutes an “effect.” As another example, the difference between legitimate and illegitimate activities is unclear among like-minded scholars and policymakers. Among rivals, that difference is drastic.

AI may be able to inspect complex systems and explain them in ways that are easier to understand or in terminology that is more familiar to its user. This is already a strength among status quo AI systems. The lack of precise and agreed-upon terms may less problematic if AI can understand what one person is likely to mean when they use a term and help translate that into appropriate terminology for another reader.

It may be that some confusion is irreducible or intentional though. Uncertainty about what specifically constitutes an “effect” allows aggressors the flexibility to conduct operations while denying that they caused “effects.” Imprecision also provides victims more ability to claim that attacks were illegitimate. Imprecision may persist because of its benefits even if AI provided some means of reducing or eliminating it.

AI may also be used to create software, libraries, and networks in ways that increases confusion. AI is famously difficult for humans to interpret and so too might be its creations, especially if they are written in ways or languages that are more convenient for machines than for humans. The internet and its components could also become more opaque to humans if it evolves too quickly or in ways that humans did not direct.

\subsection{Insufficient and Competing Authorities}
The bureaucracy of cybersecurity can be significant. For attackers, bureaucracy can limit the number or types of organizations that are authorized to perform cyber attacks, the types of attacks they are authorized to perform, and the extent of attacks they are allowed to perform. On defense, even humans need to request permission before performing many defensive actions such as killing processes, removing users, or taking a service offline. Other defensive authorities relate to which government agencies are responsible for which systems, how much visibility defensive companies are granted into the systems they defend, and how far forward a team is allowed to defend.

Delegating authority to AI systems will continue to be a challenge regardless of the AI's technical capability. Decisions about which software is allowed in an organization or whether to air gap a critical system can be substantial business decisions that humans may prefer not to relinquish to AI. More tactically, AI agents may not be given full autonomy to do whatever they are capable of in either a defensive or offensive context because of reliability concerns or legal liability for instance. Even without human involvement, AI systems would likely still need to confer with other defensive or offensive agents to assess how their actions will impact those other agents. And humans will likely want to retain bureaucratic control or guidance over even superhuman AI agents. 

It may be that conflicting interests among stakeholders are fundamentally unresolvable such that there is no bureaucratic structure of authorities that satisfies all stakeholders. Some stakeholders may prefer to maximize performance while others would sacrifice performance for security or reliability. Competing and insufficient authorities may be a fact of life that no amount of intelligence in an AI can resolve.

\subsection{Difficult Command and Control}
Command and control can be challenging for attackers where rogue operators can act outside of leadership visibility \citep{libicki2009cyberdeterrence}. Command and control is also technically difficult as cyber attacks typically occur within the defender’s networks so that too much communication with the attacking agent or implant could alert defenders. That communication may also need to run through obfuscation networks to protect the attacker’s identity and digital location. 

Reliable and independent AI agents could be used to reduce the need for remote command and control. The attack agents could make tactical, or potentially even operational or strategic decisions on their own. Independent agents on the target’s systems, or on the compromised hosts of a botnet, may be severely limited in resources such as neural network size, memory, and compute availability. Those limitations could prevent those agents from perform adequately or competitively with defensive agents that could operate more capable models on more performant hardware closer to the target. 

Even if those command and control agents have sufficient resources and are hosted on large remote servers rather than the target networks, they may still increase risks by performing poorly. Delegation of command and control to AI systems could happen for reasons of speed or scale rather than effectiveness. That could lead to less reliable decisions and more mistakes or miscalculations. 

\subsection{Heavily Classified}
Cyber tools and techniques, especially offensive ones, are often highly classified. That classification sustains a disparity between the capabilities of various nations. While all nations and organizations have similar defensive technology, if not expertise for implementing it, threat actors differ widely in the offensive technology available to them.

It is unlikely that the most capable general-purpose AI systems will be controlled by governments that can classify them. It is more likely that classified offensive tools, techniques, or datasets could be important for applying general-purpose AI systems to offensive cybersecurity. In that case, the stratification of threat actors due to classification could remain.

There are some similarities to classification in developer efforts to control access to the most capable AI. While the data or tools are not technically classified, access to them could be restricted such as through intellectual property protections, export controls, or simply economic disparities if the technologies are costly. Controlled access could not only create stratification among threat actors, but it could also stratify defensive abilities by providing weaker or stronger defensive tools to various customers and users.

AI may also complicate classified defenses. There are protections for both introducing and removing data or software to and from classified networks, but there are more protections to prevent transfers from high-to-low than low-to-high \citep{dticDiodes}. It would be challenging to introduce an AI agent to a classified network, but autonomous attack agents may force defenders to focus more on preventing removal or manipulation of information within networks. A reliable and independent agent may be better able to navigate the classified network without command and control and better able to exfiltrate or alter data. But those attack agents may require substantial compute resource to be effective and might be difficult to hide from classified network defenders.

\subsection{Tactical Engagement is Basic Unit of Analysis}
It is easier for analysts to consider success or failure in cybersecurity at the tactical level. The impact of an individual hack is often calculated or evaluated only in isolation. That may be becoming less true over time. It may also be less true with persistent engagement, where a single threat actor and defender interact continuously. And it is less true when cyber attacks are used as enabling aspects of broader military or influence campaigns.

If tactical decisionmaking is delegated to AI systems, then humans will need to provide non-tactical guidance to those systems. In that case, human operators would need to focus more on providing that guidance. Analysts may subsequently also focus more on assessing that guidance and interest in strategic or longer term effects may increase in prominence among both cyber operators and cyber watchers. 

Even without delegating operational or strategic authority, AI could increase the rate and volume of small-scale attacks to an extent where individual attacks are less noteworthy but their strategic impact is significant. Unable to keep pace, or uninterested in commonplace hacks, analysis may shift to bigger picture effects, objectives, and trends. That behavior occurs in widely distributed vulnerabilities such as the log4shell vulnerability \citep{log4shell}. Additionally, the large number, and often individually small effects, of ransomware attacks have accumulated to large national and global effects and a shift toward strategic considerations rather than tactical engagements.

\subsection{Conflict Escalates Horizontally and Vertically Within Cyberspace but not Yet Out of It}
Responses to cyber attacks have mostly been restrained and restricted to retaliatory cyber attacks. There are exceptions such as some arrests or sanctions \citep{cyberSanctions}, but escalation to kinetic or military responses has been limited.

Escalation outside of cyber could be more common if cyber attacks are used for more damaging effects either in isolation or in aggregate. That is a possible, but not necessary, outcome of greater AI involvement. Rather than increasing severity, it may be more likely for AI to increase the number of attacks and for nations or other targets to respond to the increased volume outside of cyber. But if AI increases the ease of cyber attacks, then that would make cyber attacks an even more attractive option for retaliation, increasing the likelihood that escalation stays within the cyber domain. 

AI could also increase the risk of accidentally crossing the escalation threshold that separates cyber aggressions from non-cyber responses. That could happen by AI increasing the number of cyber attacks so that the probability of one or a few crossing the threshold increases. It could also be the result of delegating decisionmaking authority to less qualified AI agents because they are faster or more scalable.

\subsection{Cyber Conflict May Invite Escalation, Miscalculation, and Instability}
Cyber attacks occur constantly. Each attack is unique in target, method, or circumstance. Even attacking the same target twice is different the second time because the victim may decide that isolated attacks are tolerable but that repeated attacks are unacceptable.

Increasing the volume of attacks can increase the risk of escalation and miscalculation, as can delegating authority to an unfit AI system. AI is poor at understanding the human mind and may struggle to reliably anticipate human responses to its actions. That may be less true in the future if AI improves its ability to understand humans, especially if it has access to training data that is representative of its victims and their cultures or beliefs.

If an AI system is used to suggest escalation pathways, or potentially even to help aggressors or retaliators choose among them, then its designers have the chance to determine its proclivity or reluctance to escalate from cyber attack. Decisionmakers could also use the AI system to provide recommended responses to their actions before conducting those actions and expect that its suggestions would be similar to those that their adversary's AI might provide. That could reduce miscommunications and miscalculations that would otherwise lead to escalation. It would allow adversaries to toe the line while remaining short of escalation, although the model could still contain randomness so that adversaries could not be perfectly certain of the response. The ability to toe the line could be treacherous.

Practically though, manipulability of AI models and their unreliability in the presence of intentionally subversive behavior make it unwise to cede too much decision-making authority to AI even among actors who fear miscommunication above other risks. AI would need to be extremely capable to contribute significantly to strategic decisions given that groups of expert humans are available for strategic decision-making.

\subsection{Internet Has Common Mode Failures}
Much of the technology across the internet is shared. Companies sell their wares and services to many other companies and organizations. Further, the same protocols and procedures for communication are required for devices to interoperate effectively. And open codebases and libraries are integrated across systems and software to extents that are difficult to observe.

AI may further centralize techniques and approaches, possibly creating more common mode failures. If the same, or similar, AI systems write most software, then even software that is developed at different organizations may become more similar. Vulnerabilities that come from a particular coding practice or preference exist wherever an author used that practice or preference. If one or a few distinct AI authors write an increased fraction of the code across the internet then those practices or preferences will be shared more widely.

A similar risk could arise if one or a few AI systems provide defensive services. A weakness in the techniques that the AI defender relies on, or the settings it prefers, could expose many targets simultaneously.

If AI creates more components, drawing on more diverse resources such as those designed in languages that human developers would not speak, the cyber environment could also become more interconnected and interdependent.

Each of these ways that AI could increase common mode failures should be considered in the context that cybersecurity already draws heavily on shared code and an oligopoly of defensive service providers. There are a few major operating system developers, many companies are hosted by the same few cloud services, and there are a few primary antivirus providers who work closely with each other. That AI models or systems could be simply instructed to provide variability in its activities could be enough to reduce rather than exacerbate common mode failures.

\subsection{Large Number of Devices Grants Advantage}
Having a large number of devices allows an attacker to increase the scale of their attack, especially in attacks that directly rely on volume such as DDoS attacks. Scale can also be an advantage in stealth while defenders try to identify the source of an attack, or in remaining effective while defenders disable attacking devices. This premise is easy to overstate though as many particularly damaging attacks require no more than a laptop and because having many devices on defense increases the attack surface.

Currently, AI is resource intensive, demanding many devices in order to perform its calculations. That may be less true in the future if efficiency continues to increase rapidly. But if AI systems continue to be resource intensive to operate, then the number of available devices to operate capable AI systems could be a significant determinant of the number, speed, or capability of AI agents or systems. The competition between attackers and defenders could be driven by the number of devices they have to power their offensive or defensive campaigns.

For attackers, taking advantage of the scale of AI will likely require even more devices that are widely distributed in order to have multiple attacking agents. If too many attacks come from a single device then defenders will be easily alerted and can take actions to block those connections or disrupt those devices.

\subsection{Cyberspace Advantages Those that Operate Persistently}
Some say that attackers benefit more from operating persistently than sporadically. To some extent, this is at odds with the premise that surprise is important. The argument is that “the primary [cyber faits accomplis] and secondary [direct cyber engagement] behaviors of States in and through the cyber strategic environment . . . are consequences of a structural imperative to persist and of a structurally derived strategic incentive to pursue gains through cyber exploitation short of armed-attack equivalence,” \citep{fischerkeller2022cyber}. Put another way, cyber attackers prefer to cause a death by a thousand paper cuts that each remain below the threshold of war \citep{harknett2024america}. Further, intruders can gather more information by persistently engaging an adversary, and intruders may be better prepared to take spontaneous action if they have already conducted some aspects of attack such as initial access and reconnaissance. 

AI will not decrease the value of intelligence gathering, so nation-state attackers will still desire persistent engagement. If AI decreases the timeline for conducting an end-to-end attack, then persistent engagement may be less necessary among attackers whose goals are to maintain the option to attack a target in the future. But criminals, as well as some nation-states and terrorists, often have little need for persistence, prefering to cause effects as soon as possible. AI is unlikely to change those motivations.

If vulnerabilities or novel techniques are available to attackers but in limited quantities, then exhausting them to maintain or re-establish persistence may be unwise. Attackers could prefer not to persistently engage for risk of burning capabilities that they would like to retain. Perhaps more likely though, if vulnerabilities or techniques are limited, then attackers could worry that defenders would independently discover the vulnerability or technique. An attacker may also worry that other attackers would discover, use, and burn the new vulnerability or technique first. So the incentive to use a capability before it is burned might further incentivize persistent engagement.

If AI-enabled attack and defense occurs quickly and defensive successes are relatively long-lasting, then persistent engagement could become less desirable for attackers. For example, defensive success could involve disrupting attacker infrastructure and reestablishing that infrastructure could be expensive or time-consuming. In that case, an increased rate of attack and defense cycles could be frustrating or costly for attackers and could lead to an overall reduced fraction of time where attackers have access. On the other hand, frequent but non-persistent access is still persistent engagement.

\section{AI's Influence on Cyber}
\label{influenceSection}
This section collects the various lines of reasoning that came up in section \ref{asymmetriesSection} while evaluating AI's prospective impacts on the arguments for cyber's offensive and defensive advantages and in section \ref{characterSection} evaluating AI's prospective impacts on the character of cyber conflict. These are the lines of reasoning for how AI might affect cyber in the future. Many of those lines of reasoning were repeated or similar across several of the arguments but each is included only once each in this section.

After collecting the full set of ways we assessed that AI might affect cyber, we tried to group them by similarity. We present them here in those groups which we have labeled as: changes to the digital ecoystem, hardening digital environments, tactical aspects of digital engagements, incentives and opportunities, and strategic effect on conflict and crisis.

\subsection{Changes to the Digital Ecosystem}
In addition to providing tools and capabilities to both attackers and defenders, AI is already altering the digital environment as AI is integrated into more systems and services and as it is used more by developers \citep{jetBrains2024Ecosystem}. This section outlines ways that we expect AI could continue to change what it is that attackers attack and defenders protect, as well as where those attacks and defense occur.

\begin{center}
\begin{table}[h!]
\begin{tabular}{|p{0.1\textwidth}|p{0.7\textwidth}|}
 \hline
 \multicolumn{2}{|l|}{\textbf{Digital Ecosystem}} \\ \hline 
 \multicolumn{2}{|l|}{Impact on Coding, Design, and Services} \\ \cline{2-2}
  & Produce more code and services to protect \\ \cline{2-2} 
  & Increase complexity of code, services, and architectures   \\ \cline{2-2}
  & Help make sense of complex systems  \\ \cline{2-2}
  & Increase or decrease systematic vulnerabilities  \\ \cline{2-2}
  & Increase or decrease systemic vulnerabilities  \\ \hline
 \multicolumn{2}{|l|}{Integration of AI Components} \\ \cline{2-2}
  & Incorporate AI components that are vulnerable  \\ \cline{2-2}
  & Remove human weak links  \\ \cline{2-2}
  & Remove manual control or backups and deskill humans  \\ \cline{2-2}
  & Produce new targets in the form of AI models  \\ \hline
 \multicolumn{2}{|l|}{Structure of Networks and the Internet} \\ \cline{2-2}
  & Rearrange networks toward a hub-and-spoke architecture  \\ \cline{2-2}
  & Enable more isolation and air-gapping  \\ \hline
\end{tabular}
\label{tableDigitalEcosystem}
\end{table}
\end{center}

\subsubsection*{AI’s Impact on Coding, Design, and Defense}
As a first order effect, as AI accelerates software development, defenders will find themselves with more code and services to protect. In addition to a greater number of digital elements, the code and services could be more complex or opaque because they were made by AI. AI may design its components in ways or languages that are less convenient for humans. On the other hand, AI may be helpful in analyzing and explaining complex code bases or systems and services in ways that help humans quickly understand them.

The rate that AI systems can provide or adapt services and code may be limited by the need to maintain usability. Changes that are too rapid may cause more problems than they are worth, especially if AI use leads to a more complex and interrelated digital ecosystem.

AI may also affect the degree of shared and repeated vulnerabilities across the digital ecosystem. Vulnerabilities that affect broad systems and those that systematically recur are already prevelant today with many dominant service or product providers, extensive code sharing and library use, and common defensive services such as antivirus or monitoring and scanning. A few dominant AI products, or similar AI products that share the same training data, may increase these systematic and systemic risks, but AI may be well-equipped to reduce them instead. A single AI developer or defender could be instructed to incorporate randomness or vary its approaches. Variability in outputs is a parameter that AI developers can easily adjust up or down. 

\subsubsection*{Integration of AI Components}
The push to incorporate AI models and agents into more systems and services will also be a significant change to the digital ecosystem. AI systems, including AI for cyber defense, can be vulnerable to deception and manipulation. Defenders will need to manage systems built with more vulnerable components while using AI defensive tools that are likely to be vulnerable themselves. The defensive challenge will grow as the human components of targeted systems or organizations are replaced by AI agents that can be attacked digitally. Humans may be a common weak link in cybersecurity, but they are difficult to control digitally.

Integration of AI could also decrease resilience to attack. Over time, AI systems could replace manual controls that are critical in the event of a failure or cyber attack. For example, autonomous vehicles with no steering wheel, or industrial control systems with no console, cannot be operated by humans in the event that an AI operator is compromised. Even if the interfaces remain, humans may lose the skills to operate them \citep{lee2025impact}. 

AI may also create new targets. The objective of AI is to compile vast amounts of information or capability into a compact digital form. AI models that have subsumed large amounts of classified reporting would be far easier to steal or leak than each of the individual reports from their various silos. And AI models that track their owners’ statements, behaviors, and preferences make for lucrative or damning cyber targets.

\subsubsection*{The Structure of Networks and the Internet}
AI may alter the architecture of the internet as well. If a few AI developers control the market and AI becomes the dominant route to information, then the internet could reshape from its current “scale-free” structure to more of a “hub-and-spoke” model. Today, information flows between various users and various information providers. In an alternative future, information could flow primarily from information generators to the AI companies. Users would then reach out to AI companies for access to the models that would interact with information providers on the users' behalf. A hub-and-spoke internet would have more single-point-failure risks but would also let defenders focus their efforts on those critical nodes.

Even without centralization to a few core AI providers, AI adoption could restructure the internet. If AI systems contain much of what they need to perform their functions, then they require fewer network connections, or possibly no network connections. That could lead to more air-gapped systems. The partnership between Apple and OpenAI may presage this approach. Embedded AI on the iPhone handles as much as possible without a network connection and only reaches out to OpenAI when necessary. OpenAI can then reach out to other information providers as necessary to return information to the iPhone user. This increased isolation makes AI-enabled devices easier to defend. That is, unless the AI is used to collect information from a more diverse set of sources than humans do today. In that case the opposite could be true.

Attackers may also become more reliant on distributed networks and deniable infrastructure across the internet to conduct their attacks. If the scale of attacks increases, then more attack origins will likely be required to avoid defenses that block or disrupt those attacks. The production and maintenance of that attack infrastructure could become a bottleneck if it does not accelerate as much as other aspects of attack or if an increased number of attacks allows defenders to more easily trace back and disrupt that infrastructure.

\subsection{Hardening Digital Environments}
Most cyber defense happens long before an attack. Cyber defense occurs in the design of digital systems and their functionality. It occurs during coding and testing before a product is released. And it occurs during risk assessments and configuration management such as decisions about access controls, multi-factor authentication, or firewall rules.

\begin{center}
\begin{table}[h!]
\begin{tabular}{|p{0.1\textwidth}|p{0.7\textwidth}|}
 \hline
 \multicolumn{2}{|l|}{\textbf{Hardening Digital Environments}} \\ \hline
 \multicolumn{2}{|l|}{Codebases} \\ \cline{2-2}
  & Write code that is more or less secure  \\ \cline{2-2}  
  & Refactor insecure code \\ \hline 
\multicolumn{2}{|l|}{Configurations} \\ \cline{2-2}
  & Implement standard defenses and best practices  \\ \hline  
\multicolumn{2}{|l|}{Updates and Patches} \\ \cline{2-2}
  & Find vulnerabilities  \\ \cline{2-2} 
  & Test and assess patches and updates  \\ \cline{2-2} 
  & Generate patches more quickly  \\ \cline{2-2} 
  & Exploit more hard-to-patch vulnerabilities  \\ \hline 
\multicolumn{2}{|l|}{Risk Acceptance or Rejection} \\ \cline{2-2}
  & Assess digital inventory  \\ \cline{2-2} 
  & Brainstorm risk scenarios  \\ \cline{2-2} 
  & Red-team to assess security posture or risks  \\ \hline 
\end{tabular}
\label{tableHardeningEnvironment}
\end{table}
\end{center}

\subsubsection*{Codebases}
There is a major shortfall in the number of cyber professionals across all industries. There is also a shortfall in the cyber defense budgets of small organizations, which often include critical infrastructure providers. AI’s ability to uplift defenders or to execute defensive tasks independently can help close the security holes that are due to talent shortfalls.
This hardening will likely help smaller, lower-profile organizations and contributors more than major providers who already employ highly skilled teams to securely develop, test, and update their products. Those skilled teams cannot address every flaw, but they can quickly fix high-priority ones. 

AI may also provide more benefit in securing code from smaller contributors, especially open source code, that is either critical in its own right or in that it is incorporated into important products. Assessing and securing that code is not an especially difficult task aside from the size of the codebase and the monotony of the work. With relatively modest advances, AI may be able to harden vast codebases–work that humans have neither the time nor interest to do.

\subsubsection*{Configurations}
Beyond the codebase, configuring an organization’s defenses can be a substantial task. The National Institute of Standards and Technology’s 800-53 guidance for securing information systems alone is almost 500 pages long. Executing the prescribed tasks is not always difficult but it is a big job, especially for small organizations. Accelerating, or independently conducting, these best practices could close many of the security holes that have led to major breaches and effects. AI systems may also be able to implement or monitor a greater number of defensive tools, increasing the depth of the defensive gauntlet that attackers face.

\subsubsection*{Updates and Patches}
AI is likely to increase the rate of new vulnerabity discoveries \citep{googleBigSleep} and decrease the time to develop patches for them. But historically, the time to develop updates and patches has not been the primary defensive challenge, it often takes a much longer time to implement the patch or update. Those updates can interfere with user experience, can require downtime such as for device restarts, or can lead to widespread outages if the updates are faulty. There is likely more opportunity for AI to impact defense by accelerating the test and evaluation process for updates than by accelerating the development of those updates. Even without new testing tools, AI may help system administrators quickly assess how the complexities of new updates may interact with the complexities of their own unique organizations and systems.

Reliability is not the only hurdle to overcome in update implementation. Updating too frequently can interfere with usability or impact uptime. If AI makes frequent updates to a service, then the other systems that rely on that service could become unstable or inconvenient to use.  

Patches could also be slow to generate in some cases. Hardware vulnerabilities, for example, might not always have an easy patch. And if a hardware vulnerability can be addressed by a software patch, it might come with performance or usability sacrifices. The system administrator would have to accept or reject those sacrifices and they might choose some level of vulnerability in exchange for performance.

\subsubsection*{Risk Acceptance or Rejection}
Perhaps more importantly than any tactical defense, defenders choose the level of risk they are willing to accept. If they are dissatisfied with the level of security that their defenses provide, then they can choose to disconnect a system or not offer a service, function, or feature. A major challenge for defenders has been in accurately assessing that risk in order to make those decisions.

AI can help assess risk in several ways. It may help complex organizations understand their operations, supply chains, and digital inventory, including shadow IT. AI can help brainstorm risk scenarios to better cover the set of possible risk scenarios. Additionally, proliferation of offensive cyber agents would allow defenders to conduct better red-team evaluations. In a future scenario where offensive agents are widely available and capable, defenders would be able to use those same tools to accurately evaluate their risk. They could conduct those evaluations before making their products and services available and adjust the risk to meet their tolerance.

\subsection{Tactical Aspects of Digital Engagements}
Digital engagements can be near instantaneous in cases such as SQL injection attacks where a single query to a vulnerable service can leak information. More often, it is a protracted process of offensive and defensive tactical actions and responses. In the case of persistent engagement, it is a continual process. And defenders may face several simultaneous attackers or many instantiations of a single attacker such as during a Distributed Denial of Service attack.

\begin{center}
\begin{table}[h!]
\begin{tabular}{|p{0.1\textwidth}|p{0.7\textwidth}|}
 \hline
 \multicolumn{2}{|l|}{\textbf{Tactical Aspects of Digital Engagements}} \\ \hline
 \multicolumn{2}{|l|}{Scale Increases} \\ \cline{2-2}
  & Increase the number of attacks  \\ \cline{2-2}
  & Increase the number of defenses  \\ \cline{2-2}
  & Inspect more alerts  \\ \hline 
 \multicolumn{2}{|l|}{Speed Increases} \\ \cline{2-2}
  & Accelerate offensive tactics  \\ \cline{2-2} 
  & Accelerate defensive tactics  \\ \hline 
 \multicolumn{2}{|l|}{Availability of Resources} \\ \cline{2-2}
  & Provide less capability to offense than defense  \\ \hline 
 \multicolumn{2}{|l|}{Delegation} \\ \cline{2-2}
  & Reduce command and control  \\ \cline{2-2} 
  & Increase or decrease odds of errors and accidents  \\ \hline 
\end{tabular}
\label{tableTacticalAspects}
\end{table}
\end{center}

\subsubsection*{Volume of Attacks}
There have been autonomous attacks almost as long as there have been digital belligerents, so the scale of attacking bots may not increase as a result of AI \citep{orman2003morris}. But more advanced AI may mean that these independent autonomous attackers act more independently and could require more individualized defensive attention than worms or conventional botnets that are mostly copies of a single piece of software. More reliable and independent attack could further reduce barriers to entry, allowing new threat actors to conduct cyber offensives. For instance, terrorists may have only conducted few cyber operations because they have limited access to the expertise that more advanced AI may be able to provide. 

At the same time, advances in AI provide expertise to defense. Relatively modest levels of AI advancement could be used to help meet the national and global shortfall in cyber defenders. Those defenders may also be able to implement more defenses that operate more independently. Those defenses could be set to trigger or alert more aggressively if AI can help investigate those alerts, reducing the burden on defenders who are overwhelmed from investigating false positives. 

\subsubsection*{Speed}
An often touted benefit of AI is its speed, but not all aspects of cyber engagements can be accelerated by AI and there are limits to how much some aspects can be accelerated. AI models tend to be computation-intensive and slow compared to many other digital processes which are already fast and can be difficult to further accelerate. For example, data transfers, directory scans, or hard drive encryptions would not be easy to further accelerate by AI. 

There are hard limits to how much faster many of these things could occur based on bandwidth, circuitry, or information theory for example. And defense can choose to further slow certain processes such as by requiring human approvals after a password lockout or before encrypting or deleting important files. Even in some cases where the offense could accelerate some processes, they may choose not to in order to avoid being flagged for malicious activity. 

Some elements of digital engagements can be accelerated though. There are tasks that are currently slow such as reverse engineering a binary or vulnerability discovery. If slow tasks are bottlenecks in an operation, then accelerating those tasks could provide substantial overall speedups. 

If some aspects of digital engagements are accelerated by AI, other aspects that are harder to accelerate may become more important. Cyberspace makes distant geographies accessible, but even the speed of light may be too slow and can induce meaningful delays, especially for attackers who are geographically removed from their targets. Attackers may also need time to re-establish infrastructure and deniable accounts that defenders might destroy or block during an engagement.

\subsubsection*{Availability of Resources}
Perhaps more importantly than geographic delays during engagements, the defense will often be able to use more capable AI systems and agents. Defenders can allocate whatever resources they choose to their AI defenders, and defenders can pass as much information to and from those AI agents as they wish. Attackers will often be restricted to using limited resources on the victim systems that will not alert defenders. Alternatively, attackers can pass information to and from remote servers, but they still need to avoid alerting defenders with those data transfers. 

That does not necessarily mean that defense will be easier than attack because the tasks that each conducts during an engagement are so different. Cracking a password is different from creating a password, and performing a malicious edit to code is different from identifying an edit as malicious. The preferred tactics of digital engagements will likely shift as AI aids in some tasks and does not advance others.

\subsubsection*{Delegation}
Tactical, or even operational, decisions may be delegated to AI cyber agents because of various benefits such as minimizing command and control, increased speed, or simply because there are too many agents for humans to manage individually. The ability for attack agents to operate without command and control could allow them to be more useful against hard targets such as classified networks or air-gapped systems.

If authority is delegated to those agents for reasons other than their reliability, they may not be well-equipped to make those decisions, which could result in more mistakes. That is true for both offensive and defensive agents, but offensive organizations may be more risk-tolerant.

On offense, those mistakes might be accidental escalation or collateral damage. On defense, the agents may kill processes, disconnect systems, or wipe endpoints. Those could be very disruptive to users or to organizations that are trying to provide reliable services. If AI-enabled defenders are also vulnerable to manipulation then that risk could be especially problematic. The attackers may not need to initiate effects themselves if they can induce defensive agents to cause those disruptions or damages instead.

\subsection{Incentives and Opportunities}
Offensive cyber is often thought of as an asymmetric and accessible form of aggression. It requires less investment, population, and industrial base than other domains. But the cyber threat environment is still widely differentiated. Script kiddies, terrorists, and hacktivists are far more limited in capability than nation-states. Even  among and within nation-states, the capabilities of various teams differ greatly. 

\begin{center}
\begin{table}[h!]
\begin{tabular}{|p{0.1\textwidth}|p{0.7\textwidth}|}
 \hline
 \multicolumn{2}{|l|}{\textbf{Incentives and Opportunity}} \\ \hline
 \multicolumn{2}{|l|}{Availability to Attackers} \\ \cline{2-2}
  & Lower the barrier of entry for aspiring attackers  \\ \hline 
 \multicolumn{2}{|l|}{Attribution} \\ \cline{2-2}
  & Analyze vast and disparate information sources  \\ \cline{2-2} 
  & Reduce human involvement in attacks  \\ \cline{2-2} 
  & Perform false flag operations  \\ \hline 
\end{tabular}
\label{tableIncentives}
\end{table}
\end{center}

\subsubsection*{Availability to Attackers}
The more that AI proliferates and uplifts offensive capabilities, the more capable lower-level threat actors become. Attack tools and scripts are already widely available to individuals, terrorists, or criminals. AI will progressively make those tools more formidable and easier to use \citep{anthropic2025}. Whether those tools, and the attackers they enable, become more effective and threatening depends on whether defenses harden more than attackers progress. 

The scalability and availability of defensive AI agents may allow defenders to better protect against known threats. In that case, the widespread availability of known attack tools to exploit known vulnerabilities may become less of a problem in the future, even if those tools are AI-enabled. If the AI-enabled attack tools are useful for discovering novel vulnerabilities or techniques, then defenses may struggle. Such a capability would be the equivalent of what only the most advanced nation-state threats possess today but perhaps at larger scales. If it became widespread, that would reduce the disparity between the most and least capable threat actors.

In some scenarios, the gap between threat actors could remain significant. As an example, if capable offensive agents are only available from a few providers, then those providers may be able to restrict access. Even if rival countries have competitive offerings, they may agree to prohibit use by terrorists or some criminal organizations. 

As another example, offensive agents may require sensitive or classified tools, datasets, or expertise that are only available to a few organizations. In that case, offensive AI agents may be capable but not widely available. Those tools might then need to be deployed in restricted ways. If they were deployed in unprotected networks or target environments, then adversaries or various other threat actors could most likely replicate and repurpose those tools or aspects of them. 

\subsubsection*{Attribution}
In some cyber attacks, the aggressor does not worry about being identified, or may even want to claim credit for their attack. Often though, attribution matters for deterring or punishing cyber attackers, and AI could accelerate the attribution process to be nearer to incident timelines or make attribution more difficult.

AI may be able to make sense of more disparate pieces of information to connect the dots or make a stronger case for or against a certain suspect. AI excels at pattern matching and may be able to identify similarities between attacks that humans would miss. But if AI performs a greater share of an attack, there would be fewer clues from its human perpetrators. If the same AI tools are used by different attackers then those attackers would be harder to distinguish. Further, AI may be adept at conducting false flag operations. For example, it could be instructed to write scripts in the style of other known hacking groups.

But attribution does not exclusively rely on the technical aspects of attack that AI would affect. Other clues such as tensions and motivations would remain, as might a trail of money or exfiltrated data. And the location or ownership of command and control servers could still be revealing, as long as the agents are still being commanded and controlled.

\subsection{Strategic Effect on Conflict and Crisis}
Cyber attacks have had limited strategic effect. They have certainly caused extensive financial damage, disruption, and even death, but as compared to other domains, cyber has provided less strategic influence and has directly achieved fewer strategic objectives. But that may be primarily because the cyber developed as a domain during a period where the strategic interests of major rivals was to avoid escalation. 

\begin{center}
\begin{table}[h!]
\begin{tabular}{|p{0.1\textwidth}|p{0.7\textwidth}|}
 \hline
 \multicolumn{2}{|l|}{\textbf{Strategic Effects on Conflict and Crisis}} \\ \hline
 \multicolumn{2}{|l|}{Decision Processes} \\ \cline{2-2}
  & Increase decision speed  \\ \cline{2-2} 
  & Improve or reduce decision quality  \\ \hline 
 \multicolumn{2}{|l|}{Strategic Value of Cyber} \\ \cline{2-2}
  & Accelerate operations or tactics to meet strategic need  \\ \cline{2-2} 
  & Force disconnection or hardening of high-risk systems  \\ \hline 
 \multicolumn{2}{|l|}{Certainty} \\ \cline{2-2}
  & Provide certainty that at least one attack will succeed  \\ \cline{2-2} 
  & Increase risk of accidental escalation or collateral damage  \\ \hline 
 \multicolumn{2}{|l|}{Miscommunication and Miscalculation} \\ \cline{2-2}
  & Improve or impair understanding of likely adversary behavior and responses  \\ \cline{2-2} 
  & Reveal offensive intentions in forensic analysis  \\ \hline 
 \multicolumn{2}{|l|}{Use It or Lose It} \\ \cline{2-2}
  & Increase transience of cyber capabilities or vulnerabilities  \\ \cline{2-2} 
  & Increase availability and replaceability of capabilities or vulnerabilities \\ \cline{2-2} 
  & Increase or decrease ability to achieve offensive objectives on-demand \\ \hline
\end{tabular}
\label{tableStrategic}
\end{table}
\end{center}

\subsubsection*{Decision Processes}
Strategic decisions about whether and how to use offensive cyber capabilities or about substantially changing defensive postures can be slow. They are often multi-faceted decisions with many different simultaneous objectives to achieve and risks. The heads of various departments of a government or organization may have perspectives that are different from each other based on the needs of their departments, and those perspectives may be complex and unstated. AI may be able to accelerate or improve these decisions but there are reasons to think its effect may be limited.

Humans will probably want to, or be required to, remain involved in the decision-making process even if AI becomes extremely capable. AI may be able to accelerate much of the process but human bottlenecks may remain when others have been removed, limiting how much the process as a whole can be accelerated. And it may be that no amount of intelligence can remove conflict and tradeoffs from strategic decision-making in ways that improve the quality of outcomes for all stakeholders.

\subsubsection*{Strategic Value of Cyber}
Part of the reason for cyber’s historically lower strategic value may be that defenders do not accept substantial risk. If a service provider cannot secure a service sufficiently well, then they may choose not to offer that service. For instance, it would be convenient and efficient to vote remotely over the internet but that would be difficult to protect. Defenders will continue to have the option to not offer certain services or functionality, but if offense becomes more capable, then defenders may feel the need to discontinue existing services even in the absence of any actual cyber attack. In that way, a change in offense-defense balance could have strategic impacts even without any actual attacks.

Another possible explanation for cyber’s limited historical strategic value is that despite its tactical speed, it can be operationally slow. Missile strikes can happen in minutes or hours, and ground invasions in days or weeks. Cyber operations often require substantial preparations and tailored development matching the attack to its target. AI may accelerate these processes, making cyber options more available to decision-makers on the timescales they need.

\subsubsection*{Certainty}
Cyber offensives are also highly uncertain, making them difficult for leaders to rely on for operations of strategic importance. If offensive AI agents increase the number of attacks that can be performed in a given day or week, either because they are performed faster or by a larger number of independent agents, that increases the likelihood that at least one of those attacks will succeed. Decision-makers can be more certain that their operations will achieve their strategic objectives.

At the same time, a larger number of attacks in a given period of time increases the likelihood of mistakes. It can increase the chances of unintended escalation or of collateral effects. While pursuing more strategic objectives, these accidents could be more impactful than they have been in the past.

\subsubsection*{Miscommunication and Miscalculation}
In addition to unintended consequences that are either escalatory or that victimize unintended targets, intended effects can have unintended consequences. AI may be particularly weak at understanding human responses and may make or advise decisions poorly. That is particularly likely if the model or agent is trained on limited data or on data from only the aggressor’s culture. 

But if appropriately trained, AI may be able to suggest victim redlines or responses that human aggressors would not anticipate. If the victim decision-makers are being advised by AI, it may even be possible for the aggressor to acquire a similar AI system and evaluate the responses that system would advocate for prior to conducting those aggressions. 

Responding to aggression also requires some understanding of the extent and intent of the aggression that is not always clear. Cyber analysts debate whether foreign malware in critical infrastructure is legitimate espionage or illicit prepositioning for destructive attacks. Captured offensive AI agents may not only reveal what they are capable of doing as analysis of offensive tools reveals today, they may also reveal more about what they intend to do. Security analysts may be able to test captured AI agents to determine the conditions under which they would decide to perform damaging or escalatory actions. Those results may be more definitive in forensic analysis following attack than in anticipation of an attack if the human aggressors could easily update benign agents to be more aggressive.

\subsubsection*{Use It or Lose It}
Decision-makers can also be pressed to launch cyber attacks due to fear that those attacks will not be available in the future. The transience of cyber tools, accesses, and exploits may already drive threat actors to be more aggressive. They may race to use exploits as soon as they are available or be more aggressive earlier in a crisis or conflict before defenses can be hardened. 

If AI shortens the expected time until a vulnerability is discovered by defenders, revealed by another attacker, or simply removed as part of normal software churn, then attackers may feel even more pressure to act quickly and aggressively. On the other hand, if the AI technology that makes vulnerabilities faster for defenders to discover also makes them more plentiful to attackers then that pressure could be relieved.

\section{Conclusions}
There is no single answer to the question of whether AI will provide offense or defense the advantage. There are too many ways to consider benefits, such as cost, destruction, coercion, or information transfer. And there are too many aspects of cyber conflict and competition that will be affected in opposing ways.  

AI will most likely lead to a larger number of more complex products to defend that may incorporate AI components that are more vulnerable than typical digital components while decreasing the resilience that humans provide during and following cyber attacks. But AI use may also reshape the internet and connectivity, hardening targets by reducing their need for connectivity or by centralizing information flows to and from a few key nodes. 

AI is likely to further harden the broader digital ecosystem if it advances to a point where it can independently review code or organizational practices for security vulnerabilities and adapt them to adhere to best practices. Those tasks are likely to be data-rich in ways that may be amenable to AI training. If however, AI accelerates vulnerability discovery, particularly for difficult-to-patch vulnerabilities, defenders may struggle to keep up even if AI can help develop patches quickly. In that case, AI would need to help design less vulnerable systems to maintain defenses.

The speed and volume of attack and defense are already high without AI so additional intelligence may produce less change than some anticipate. But there are tactical aspects of cyber campaigns that could be accelerated by AI such as vulnerability discovery or some aspects of reconnaissance. Where those are bottlenecks, then overall speeds could increase. Perhaps more impactful than speed or volume, the delegation of tactical decisions could increase the variety of attacks that defenders face while increasing the risk of accidents. That is especially true if the delegation is motivated by speed or volume rather than reliability. Accidents could come from either offensive agents or defensive agents.

At a more strategic level, AI may lower the barrier of entry to historically less capable threat actors such as terrorists, but those actors may be less capable of conducting attacks if AI simultaneously helps harden defenses. For more traditional threat actors, accelerating cyber operations or increasing the number of independent attacks that can be attempted may make cyber a more viable option. If AI makes offense easy, then aggressors may have less need to conduct attacks, the threat alone would suffice and defenders would reduce or avoid services or operations that would be catastrophic to attack. Reduction or elimination of those services could be catastrophic enough without attacks. If on the other hand, AI helps harden targets, then there may also be increased pressure to use offensive capabilities quickly before defenders secure against them, but hardened defenses would make attacker successes rare and limited.

\section*{Authors}
Andrew Lohn is a Senior Fellow working on the CyberAI Project at the Center for Security and Emerging Technology.

\section*{Acknowledgments}
For iterative feedback and support, we are grateful to John Bansemer, Rebecca Gelles, Colin Shea-Blymer, Kyle Miller, Jessica Ji, and Chris Rohlf. We would like to thank Alexandra Fall for formatting and proofreading. And we would also like to thank Jason Healey, Bruce Schneier, and Brandon Valeriano for helpful discussions and correspondence.

\section*{Funding}
This work was supported by a grant from the United Kingdom's Department for Science, Innovation, and Technology. 

\section*{Copyright}
© 2025 by the Center for Security and Emerging Technology. This work is licensed under a Creative Commons Attribution-Non Commercial 4.0 International License.
To view a copy of this license, visit https://creativecommons.org/licenses/by-nc/4.0/.


\bibliography{references}

\end{document}